\documentclass[10pt,twocolumn,twoside]{IEEEtran}
\usepackage{amsmath}
\usepackage{amssymb}

\usepackage{graphicx}
\usepackage{epstopdf}
\usepackage{amsmath}
\usepackage{array}
\newcommand{\PreserveBackslash}[1]{\let\temp=\\#1\let\\=\temp}
\newcolumntype{C}[1]{>{\PreserveBackslash\centering}p{#1}}
\newcolumntype{R}[1]{>{\PreserveBackslash\raggedleft}p{#1}}
\newcolumntype{L}[1]{>{\PreserveBackslash\raggedright}p{#1}}
\usepackage[usenames]{color}
\usepackage{colortbl,booktabs}
\usepackage{tabularx}
\usepackage{multicol}
\usepackage{booktabs}
\usepackage[Symbol]{upgreek}
\usepackage{bm}
\usepackage{cite}
\usepackage{balance}
\usepackage[linesnumbered,ruled]{algorithm2e}
\usepackage{mathrsfs}
\usepackage{url}
\usepackage{subfigure}
\usepackage{threeparttable}
\usepackage{amsmath}
\usepackage{dcolumn}
\usepackage{multirow}
\usepackage{booktabs}
\usepackage{amsfonts}
\usepackage{amsthm,amsmath,amssymb}
\usepackage{mathrsfs}
\usepackage{algorithmic}
\newtheorem{proposition}{Proposition}[section]
\usepackage{titlesec}

\allowdisplaybreaks[3]

\begin{document}
	\title{Bayesian Sensing for Time-Varying Channels \\ in ISAC Systems\vspace{-0 mm}}
	%
	\author{Xueyang Wang,
		Kai Wu,~\IEEEmembership{Member, IEEE},
        J. Andrew Zhang,~\IEEEmembership{Senior Member, IEEE},\\
        Shiqi Gong
        and Chengwen Xing,~\IEEEmembership{Member, IEEE}
		\thanks{

X. Wang and C. Xing are with the School of Information and
Electronics, Beijing Institute of Technology, Beijing, China (email: xywang1103@gmail.com; xingchengwen@gmail.com).

S. Gong is with the School of Cyberspace Science and Technology, Beijing Institute of Technology, Beijing, China (email: gsqyx@163.com).

Kai Wu and J. A. Zhang are with the School of Electrical and Data Engineering, University of Technology Sydney, NSW, Australia. (email: Kai.Wu@uts.edu.au; Andrew.Zhang@uts.edu.au).


}
		
	}
	\maketitle
\begin{abstract}
	
Future mobile networks are projected to support integrated sensing and communications in high-speed communication scenarios. Nevertheless, large Doppler shifts induced by time-varying channels may cause severe inter-carrier interference (ICI). Frequency domain shows the potential of reducing ISAC complexity as compared with other domains. However, parameter mismatching issue still exists for such sensing.  In this paper,  we develop a novel sensing scheme based on sparse Bayesian framework, where  the delay and Doppler estimation  problem in time-varying channels is formulated as a 3D multiple measurement-sparse signal recovery (MM-SSR) problem. We then propose a novel two-layer  variational Bayesian inference (VBI) method to decompose the 3D MM-SSR problem into two layers and estimate the Doppler in the first layer and the delay in the second layer alternatively. Subsequently, as is benefited from newly unveiled signal construction, a simplified two-stage multiple signal classification (MUSIC)-based VBI method is proposed, where the delay and the Doppler are estimated by MUSIC and VBI, respectively. Additionally, the Cramér-Rao bound (CRB) of the considered sensing parameters is derived to characterize the lower bound  for  the proposed estimators. Corroborated by extensive simulation results, our proposed method can achieve improved mean square error (MSE) than its conventional counterparts and is robust against the target number and target speed, thereby validating its wide applicability and advantages over prior arts.

	\end{abstract}
	
	\IEEEpeerreviewmaketitle

\section{Introduction}\label{S1}
\IEEEPARstart{I}{ntegrated} 
 sensing and communications (ISAC) has been identified as key use case for the next generation mobile network~\cite{JCA3,JCA4,JCA5,JCA6}.
 {{\color{black}One of the ISAC system designs is communication-centric, where the sensing module is integrated into existing communication systems~\cite{JCAS1}. This design aims to extract sensing information by leveraging existing communication signals without sacrificing the communication performance~\cite{ISAC1}.}

 {{\color{black} 
 Recently, the sparse Bayesian learning (SBL) framework has attracted great attention in the field of sensing parameter estimation for ISAC systems~\cite{ISACBayes3,ISACBayes4,ISACByesa1,ISACByesa2}. 
Hu {\it et al.} proposed a two-stage scheme combining a coarse MUSIC-based estimator with a refined stochastic particle-based variational Bayesian inference (SPVBI) algorithm~\cite{ISACBayes4}. 
Gan {\it et al.} leveraged superimposed pilots and data symbols under a novel SBL model to enhance sensing performance~\cite{ISACByesa1}, while 
Chen {\it et al.} extended SBL using an adaptive pattern-coupled prior for joint angle and channel estimation~\cite{ISACByesa2}.
Tao {\it et al.} proposed a decoupled 1D SBL algorithm to reduce the complexity of 2D DoA estimation~\cite{ISACBayes3}. 
These methods demonstrate the strength of Bayesian approaches in capturing structured priors and handling complex estimation tasks. However, they are mainly developed for the time-invariant ISAC systems, which limits their applicability in high-mobility scenarios. 
In the meantime, there has been a rising needs for 5G and beyond to support high-speed user ends, such as high-speed trains and UAVs~\cite{JCA2}, etc.
Such high-mobility can cause excessively high Doppler spread, making the mainstream multi-carrier waveforms, such as  orthogonal frequency division multiplexing (OFDM), experience significant inter-carrier interference (ICI)~\cite{OFDM2,OFDM3}, complicating and degrading communications and sensing under conventional methods~\cite{OFDM,OTFSz}.
The dynamic nature of time-varying channels introduces additional challenges in estimation, requiring a more dedicated approach to address the parameter estimation problem in ISAC system.}

To cater for time-varying channels,
new modulation schemes have been proposed for high-mobility scenarios. A highly popular one is the orthogonal time frequency space (OTFS)~\cite{OTFS1,OTFS2,OTFS3}. Intuitively,  OTFS transforms the time-varying channel into a two-dimensional channel in the delay-Doppler (DD) domain, where data symbols multiplex in a near-constant channel. As a result, OTFS shows great potential for ISAC. Many studies have proposed OTFS-based ISAC frameworks for different applications based on integer delays and Dopplers.
Yuan {\it et al.} introduced an OTFS-based ISAC system and underscored the benefits of multiplexing data symbols in the DD domain, where communication channels can be inferred from sensing parameters~\cite{III}. They also proposed an OTFS-aided ISAC technique for uplink and downlink vehicular communication systems, allowing the roadside unit (RSU) to predict vehicle states based on the estimated parameters of the  vehicles~\cite{YUAN1}. Motivated by this ISAC framework, Xiang {\it et al.} developed a nonorthogonal multiple access (NOMA)-assisted ISAC network, where NOMA transmission and power allocation are optimized using the parameters estimated by ISAC~\cite{NOMA3D}. 
Similarly, Yang {\it et al.} tackled parameter association, channel estimation and signal detection as a bilinear recovery problem for sensing-aided uplink transmission, leveraging sensing parameters to enhance uplink communication~\cite{BIAMP}. However, while the delay resolution can be considered sufficient, it can be impractical to ignore the fractional Doppler due to the limited Doppler resolution~\cite{frac1,frac2}.

To address fractional Dopplers in OTFS-based ISAC transmission frameworks, several works have been proposed~\cite{SS,DFT,IIOT,REFINE}. Li {\it et al.} proposed an ISAC transmission framework based on spatially spread OTFS, utilizing angular domain discretization to simplify estimation and detection algorithms~\cite{SS}.  Moreover,  low-complexity and effective OTFS sensing and estimation methods have also been proposed, such as~\cite{DFT,IIOT,REFINE}. For instance, Wu {\it et al.} proposed a low-complexity channel estimation and data detection algorithm using conjugate gradient equalization, improving robustness to Doppler effects by applying a Fourier transform to data symbols~\cite{DFT}.  The authors in \cite{IIOT} proposed an efficient OTFS sensing by introducing a novel signal segmentation methods to construct the sensing signal matrix flexibly. Additionally, Shi {\it et al.}  presented a user state refinement method for ISAC-assisted OTFS systems, where refined angle estimates are used for joint delay and Doppler shift estimation with reduced complexity~\cite{REFINE}. These algorithms, considering fractional Dopplers, are more effective and accurate. However, due to the 2D convolution of data symbols and the DD channel in OTFS systems, the input-output relationship with fractional Doppler shifts can be complicated, making these methods computationally expensive.


A closed-form expression of the input-output relationship in the frequency domain for time-varying channels has been proposed in~\cite{ZHY1}, where OTFS is reformulated as a precoded OFDM system. This approach leverages the frequency domain to provide a more concise and computationally efficient system model compared to the DD domain, making it an attractive solution for handling time-varying channels. Building on this foundation, the authors further explored the channel representations across different domains by using discrete Fourier transform (DFT) and inverse DFT (IDFT) to bridge the gap between them~\cite{ZHY2}.  Motivated by~\cite{ZHY1}, Sun {\it et al.} proposed a sensing framework tailored for fast-fading channels, where sensing parameter estimation is performed by treating signals  within individual OFDM blocks and across multiple OFDM blocks separately~\cite{Sun}.  However, the separate estimation can lead to parameter mismatching issue when multiple targets are present. 

In this work, we develop a novel frequency-domain sensing scheme for time-varying scenarios employing the sparse Bayesian learning (SBL) framework. It allows the sensing parameters to be estimated simultaneously, hence addressing the parameter mismatch issue and significantly improving the sensing accuracy. The main contributions of this paper are summarized as follows:
\begin{itemize}
 \item {\color{black}We consider practical channel models and unify the intra- and inter-block cases for precise Doppler estimation. Specifically, we formulate the sensing parameter estimation as a 3D multiple measurement-sparse signal recovery (MM-SSR) problem, where the locations of non-zero elements in the 3D sparse matrices correspond to the integer Doppler, fractional Doppler and the delay.} By solving this MM-SSR problem, the Doppler and delay parameters can be estimated simultaneously.

 \item We propose a two-layer variational Bayesian inference (VBI) method to obtain accurate Doppler estimates for fast-fading channels. Specifically, the 3D MM-SSR problem is decomposed into two SSR problems, where the precision matrix in the first layer is calculated based on the results of the second layer. After estimating the 3D sparse matrix, we can obtain the Doppler and delay estimates according to the locations of non-zero elements.
 
 \item We also develop a simplified two-stage multiple signal classification (MUSIC)-based VBI method to reduce the complexity of two-layer iterations. Specifically, we introduce some signal construction tactics, enabling the delays to be estimated accurately using MUSIC in the first stage. We then adopt the VBI method to solve the MM-SSR problem in the second stage for a simplified Doppler estimation. Then, we derive the Cramér-Rao bound (CRB) of all parameters for time-varying sensing in the frequency domain, which provides a theoretical lower bound for the estimation of the sensing parameters.


\end{itemize}

The rest of this paper is organized as follows. In Section \ref{S2}, we present the system model of the OFDM system with multiple blocks over fast fading channels. In Section \ref{S3}, we formulate the sensing parameter estimation problem as a SBL framework and propose a two-layer VBI method to solve it. In Section \ref{S4}, we propose a simplified two-stage MUSIC-based  VBI method. In Section \ref{S5}, we derive the CRB to provide a theoretical lower bound of the sensing parameter estimation. Finally, in Section \ref{S6}, we provide our simulation results, followed by our conclusions in Section \ref{S7}.

\emph{Notations}: We use the following notations throughout this paper. We let a, $\mathbf{a}$, $\mathbf{A}$ represent a scalar, vector, and matrix, respectively; 
$(\cdot)^{{T}}$,
$(\cdot)^{{H}}$, $(\cdot)^{-1}$ and $(\cdot)^{\dagger}$ denote the transpose,
conjugate transpose, inverse and pseudo-inverse of a matrix, respectively; $\delta( \cdot )$ is the Dirac delta function; 
$a^*$ denotes the conjugate of complex numbers; $[\cdot]_N$ denotes modulo $N$ operation;  $ \propto $ indicates equality up to a multiplicative constant;  $\left <f(\mathbf{x})\right >_{p(\mathbf{x})}$ is the expectation with respect to $p(\mathbf{x})$; $q^*(\mathbf{x})$ denotes the optimal solution. Finally, $\mathbf {I}$ denotes identity matrix.

\section{System Model  and Problem Formulation}\label{S2}

In this section, we establish the system model by first deriving the input-output relationship and then providing the channel representations for the time-varying channel in the frequency domain. Accordingly, the sensing parameter estimation problem based on the known channel matrix is formulated as a MM-SSR problem. We consider a setup where there is no clock asynchrony between the transmitter and the sensing receiver. This setup could correspond to a mono-static ISAC system or a bi-static system with locked clock at the transmitter and receiver.

\subsection{Channel Representation in the Frequency Domain}

\begin{figure}[t]
	\center{\includegraphics[width=8cm]{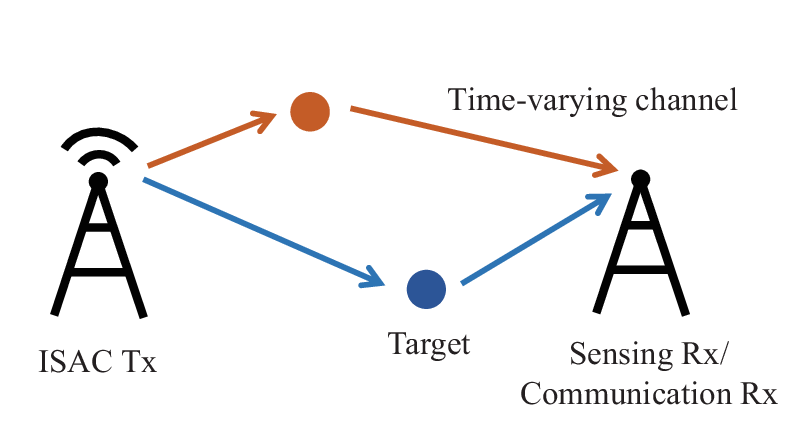}}
	\caption{The considered ISAC scenario with high-speed targets.}
	\label{diagram1}
\end{figure}
Without loss of generality, we consider an OFDM system with $K$ OFDM blocks for joint sensing and communication, where the communication receiver is also the sensing receiver as shown in Fig. 1. Assume that there are $L$ high-speed targets.  Let $h_l$, $\tau_l$ and $\nu_l$ denote the scattering  coefficient, the delay and the Doppler corresponding to the $l$-th target, respectively. 
We assume that the time duration of an OFDM block is $T = NT_0$, where $T_0$ is the time sampling interval and $N$ is the number of samples or subcarriers~\cite{Sun}. The total signal bandwidth is given by $B = 1/T_0$ while the subcarrier spacing is $f_0 = B/N$. For the $k$-th ($k=0,\ldots,K-1$) OFDM block, the delay-Doppler (DD) channel can be modeled as~\cite{Sun}
\begin{align} \label{hDD}
h_{\nu}(\tau, \nu)=\sum_{l=1}^L h_{l} e^{j 2 \pi \nu_{l} k T} g_1\left(\tau-\tau_{l}\right) g_2\left(\nu-\nu_{l}\right),
\end{align}
where $g_1(\cdot)$ and $g_2(\cdot)$ are the inverse Fourier transform (IFT) of the frequency windowing function $G_1(f)$ and Fourier transform (FT) of the time windowing function $G_2(f)$, respectively. The initial phase of the $k$-th OFDM block caused by  the target Doppler, i.e. $\nu_l$, is denoted as $e^{j 2 \pi \nu_{l} k T}$ in (\ref{hDD}).  {As is typical, we consider rectangular windowing functions for both time and frequency domains \cite{Sun}. Thus, $g_1(\cdot)$ and $g_2(\cdot)$ are both sinc functions.}

By applying the FT to $h(\tau,\nu)$ in (\ref{hDD}) with respect to the delay $\tau$, the continuous frequency-Doppler (FD) channel can be given by
\begin{align}\label{HFD}
H_{f d}(f, \nu)=G_1(f) \sum_{l=1}^L h_{l} e^{j 2 \pi \nu_{l} k T}  g_2\left(\nu-\nu_{l}\right) e^{-j 2 \pi f \tau_{l}}.
\end{align}
Let $S(f)$, $R(f)$ and $W(f)$ denote the transmitted signal, the received signal and the AWGN in the frequency domain, respectively. Then, $R(f)$ can be calculated by~\cite{ZHY1}
\begin{align} \label{Rf}
R(f)=\int_{-\infty}^{+\infty} H_{f d}\left(f^{\prime}, f-f^{\prime}\right) S\left(f^{\prime}\right) d f^{\prime}+W(f).
\end{align}

{{\color{black}Based on  (\ref{Rf}), the discrete input-output relationship in the frequency domain sampling at $f = nf_0 (n = 0,\ldots,N-1)$ and ${f^{\prime}}= mf_0 (m = 0,\ldots,N-1)$} becomes
\begin{align} \label{Rnf}
R(nf_0)=\sum_{m=0}^{N-1} H_{f d}\left(mf_0, (n-m)_Nf_0\right) S\left(mf_0\right) +W(nf_0).
\end{align}
Let $\mathbf{s}(k)$, $\mathbf{r}(k)$ and $\mathbf{H}_{fd}(k)$ denote the discrete transmitted signal, received signal and the channel matrix in the frequency domain corresponding to the $k$-th OFDM block, respectively. We can rewrite (\ref{Rnf}) in a concise form as
\begin{align}\label{rs}
\mathbf{r}(k) = \mathbf{H}_{fd}(k)\mathbf{s}(k) + \mathbf{w}(k),
\end{align}
{{\color{black}where $\mathbf{r}(k)\in \mathbb{C}^{N\times 1}$, $\mathbf{s}(k)\in \mathbb{C}^{N\times 1}$ and $\mathbf{w}(k) \in \mathbb{C}^{N\times 1}$ are obtained by stacking $R(nf_0), (n = 0,\ldots,N-1)$, $S(mf_0), (m = 0,\ldots,N-1)$ and $W(nf_0), (n = 0,\ldots,N-1)$ into column vectors, respectively.} Moreover, based on (\ref{hDD})-(\ref{rs}), the $(n,m)$-th element of $\mathbf{H}_{fd}(k)$ can be detailed as
\begin{align} \label{Hfdnm}
\left(\mathbf{H}_{f d}(k)\right)_{n, m}& = H_{f d}\left(m f_0,(n-m)_N f_0\right) \notag \\
& =G_1\left(m f_0\right) \sum_{l=1}^L h_{l} e^{j 2 \pi \nu_{l} k T} \notag \\
& \quad \quad \quad g_2\left(\left((n-m)_N f_0-\nu_{l}\right)_{N }f_0\right) e^{-j 2 \pi m f_0 \tau_{l}},
\end{align}
where $(\cdot)_N$ represents mod-N operation. 

Note that  $\mathbf{H}_{fd}$ is not diagonal if $\nu_l\neq 0$ for $l=1,\ldots,L$. Moreover, based on the expression of $\mathbf{H}_{fd}(k)$ in (\ref{Hfdnm}), we see that elements along the diagonal directions correspond to the same quantized Doppler shift while elements in the same column correspond to the same frequency.

\subsection{Problem Formulation for Estimating Sensing Parameters }
 In this work, we aim to estimate the sensing parameters $\tau_l$ and $\nu_l$ for $l = 1,\ldots,L$ {based on the channel matrix $\mathbf{H}_{fd}$.} To this end, we assume that the estimated channel matrix $\widehat{\mathbf{H}}_{fd}$ is known  for developing sensing algorithms, and will discuss the impact of {${\mathbf{H}}_{fd}$} estimation error on proposed designs. 
 
By re-aligning the elements in $\widehat{\mathbf{H}}_{fd}$, we can make elements in the same row correspond to the same quantized Doppler shift. Specifically, we can circularly shift the $m$-th column in $\widehat{\mathbf{H}}_{fd}$ upwards by $(m-1)$ positions.
Let $\widetilde{\mathbf{H}}_{fd}$ and $\widetilde{\mathbf{W}}_{fd}$ denote the re-aligned channel matrix and the re-aligned noise matrix, respectively. Then the $(n,m)$-th element of $\widetilde{\mathbf{H}}_{fd}$ for the $k$-th OFDM block becomes
\begin{align} \label{hnmk}
\tilde{h}_{f d}(n, m, k)&=H_{f d}\left(m f_0,(n)_N f_0\right) + \tilde{w}_{f d}(n, m, k) \notag \\
&=G_{1}\left(m f_{0}\right) \sum_{l=1}^{L} h_{l}e^{j 2 \pi \nu_{l} k T}  g_{2}\left(n f_{0}-\nu_{l}\right) \notag \\
& \quad \quad \quad \quad \quad \quad e^{-j 2 \pi m f_{0} \tau_{l}} +\tilde{w}_{f d}(n, m, k),
\end{align}
where the noise elements $\tilde{w}_{f d}(n, m, k)$ follow Gaussian distribution with zero mean and variance $\sigma^2$.


Estimating $\tau_l$ and $\nu_l$ for $l = 1,\ldots,L$ from $\widetilde{\mathbf{H}}_{fd}$ given in (\ref{hnmk}) can be modeled as a sparse signal recovery (SSR) problem, as illustrated next. Let $p$ and $q$ denote the indices of the delay and Doppler grids, respectively, where $p=1,\ldots,P$ and $q=1,\ldots,Q$. By considering a rectangular pulse for $G_1(mf_0)$, (\ref{hnmk}) can be rewritten as 
\begin{align} \label{hnmkCS}
\tilde{h}_{f d}(n, m, k)= \sum_{p=1}^{P} \sum_{q=1}^{Q} h_{p,q} g_{2}\left(n f_{0}-\nu_{q}\right)  e^{-j 2 \pi m f_{0} \tau_{p}} e^{j 2 \pi \nu_{q} k T}\notag \\ 
+\tilde{w}(n,m,k).
\end{align}

For a given Doppler index $n$, we can collect signals and noise over $N$ subcarriers and K OFDM blocks and arrange them into $N\times K$ matrices $\widetilde{\mathbf{H}}_{f d}(n)$ and $\widetilde{\mathbf{W}}_{f d}(n)$. Their $(m,k)$-th elements are $\widetilde{h}_{f d}(n, m, k)$ and $\widetilde{w}_{f d}(n, m, k)$, respectively. Based on (\ref{hnmkCS}),  $\widetilde{\mathbf{H}}_{f d}(n)$ can be represented in a matrix form as
\begin{align}\label{Hfd}
\widetilde{\mathbf{H}}_{f d}(n)=\overline{\mathbf{A}}_{\tau} \overline{\mathbf{D}}(n) \overline{\mathbf{A}}_{v}^{H} + \widetilde{\mathbf{W}}_{f d}(n),
\end{align}
where $\left(\overline{\mathbf{A}}_{\tau}\right)_{m, p}=e^{-j 2 \pi m f_{0} \tau_{p}}$,  $\left(\overline{\mathbf{A}}_{\nu}^{H}\right)_{q, k}=e^{j 2 \pi v_{q} k T}$ and $\overline{\mathbf{D}}(n)\in \mathbb{C}^{P\times Q}$ is obtained by stacking $h_{p,q}g_2(nf_0-\nu_p)$ for $p=1,\ldots,P, q=1,\ldots,Q$.
 Note that $\overline{\mathbf{D}}(n)\in \mathbb{C}^{P\times Q}$ is a sparse matrix with non-zero elements corresponding to the delay and the Doppler.

Furthermore, we can extend (\ref{Hfd}) to the MMV case by collecting $N$ measurements $\widetilde{\mathbf{H}}_{f d}(n)$ for $n=0,\ldots,N-1$. The MM-SSR problem can be given by
\begin{align}\label{HfdMMV}
\widetilde{\mathbf{H}}_{f d}=\overline{\mathbf{A}}_{\tau} \overline{\mathbf{D}} (\mathbf{I}_N \otimes \overline{\mathbf{A}}_{v}^{H}) + \widetilde{\mathbf{W}}_{f d},
\end{align}
where $\widetilde{\mathbf{H}}_{f d} = \left[\widetilde{\mathbf{H}}_{f d}(0),\ldots,\widetilde{\mathbf{H}}_{f d}(N-1) \right]$, $\overline{\mathbf{D}} = \left[\overline{\mathbf{D}}(0),\ldots,\overline{\mathbf{D}}(N-1)\right]$ and $\widetilde{\mathbf{W}}_{f d} = \left[\widetilde{\mathbf{W}}_{f d}(0),\ldots,\widetilde{\mathbf{W}}_{f d}(N-1) \right]$.

\emph{Remark 1}: The signal model in (\ref{hnmk}) corresponds to the inter-block sensing scenario with multiple OFDM blocks, where the accumulated phase shifts $e^{j 2 \pi \nu_{l} k T}$ are introduced in the $k$-th OFDM block. Since these phase shifts are much larger than those within a single OFDM block, the estimated Doppler range is constrained to $(-B/N/2,B/N/2]$, even though the resolution $B/N/K$ remains accurate. Furthermore, the model in (\ref{hnmk}) can be simplified to the intra-block sensing case with a single OFDM block by setting $k=0$. In this case, the estimated Doppler range can be improved to $(-B/2,B/2]$, but the resolution $B/N$ becomes relatively coarse. {\color{black}Note that the delay range to be estimated is smaller than the maximum estimable delay $\tau_{max} = 1/f_0$, which suppresses the needs for  separating the actual  delay to be estimated into integer and fractional parts of $\tau_{max}$.  As a result, there is no need to separate the delay into integer and fractional parts, and fractional delays are naturally accounted for in our model.}

Based on the signal model (\ref{hnmkCS}), we observe that the Doppler estimation can be more complicated, as $\nu_l$ appears in both $g_{2}\left(n f_{0}-\nu_{q}\right)$ and $e^{j 2 \pi \nu_{q} k T}$.  Prior art mainly divides $\nu_l$ into an integer part $n_lf_0$ and a fractional part $\xi_l$, i.e. $\nu_l = n_lf_0+\xi_l$. It has been demonstrated that the coarse Doppler estimation $n_lf_0$ can be estimated using intra-block symbols, while the fractional part $\xi_l$ can be estimated using inter-block information $e^{-j2\pi \nu_l kT}$~\cite{Sun}. However, estimating these two parts separately can lead to a challenging association and mismatch problem for different parameters.  Advanced sensing techniques, which can avoid the problem while exploiting both intra-block and inter-block signals, are desired.
Next, we propose a two-layer VBI estimation algorithm capable of simultaneously estimating the integer Doppler, fractional Doppler, and delay within the same SBL framework, thus avoiding the mismatch issue.

\section{Proposed Sensing Parameter Estimation Using Two-Layer SBL}\label{S3}

In this section, we adopt the SBL framework~\cite{bayesbase,bayesbase2,bayestutorial} and develop a two-layer VBI method to estimate the delay and the Doppler simultaneously. Specifically, we transform the sensing parameter estimation problem to two related SSR problems, which can be referred to as the first layer and the second layer. The precision matrix in the first layer is calculated based on the estimation results from the second layer. 

\subsection{Conventional SBL Framework}

In order to estimate the accurate Doppler and the delay simultaneously, we consider the inter-block signal model to estimate the fractional Doppler, where multiple OFDM symbols are considered. Specifically, we can estimate the fractional Doppler $\xi_l$ using $e^{-j2 \pi \nu_l kT}$ and the integer Doppler $n_l$ based on the non-zero $\widetilde{\mathbf{H}}_{fd}(n)$.

In the conventional SBL framework~\cite{bayesbase,bayesbase2}, the sensing signal matrix in  (\ref{Hfd}) can be rewritten in a vector form as
{{\color{black} \begin{align} \label{hn}
\widetilde{\mathbf{h}}_{fd}(n)= (\overline{\mathbf{A}}_\nu^* \otimes \overline{\mathbf{A}}_\tau) \overline{\mathbf{d}}(n) +\widetilde{\mathbf{w}}_{fd}(n),
\end{align}}
where $\widetilde{\mathbf{h}}_{fd}(n) = \operatorname{vec}(\widetilde{\mathbf{H}}_{fd}(n))\in \mathbb{C}^{NK\times 1}$, $\overline{\mathbf{d}}(n) = \operatorname{vec}(\overline{\mathbf{D}}(n))\in \mathbb{C}^{PQ\times 1}$ and $\widetilde{\mathbf{w}}_{fd}(n) = \operatorname{vec}(\widetilde{\mathbf{W}}_{fd}(n))\in \mathbb{C}^{NK\times 1}$. By collecting $N$ measurements, $\widetilde{\mathbf{h}}_{fd}(n)$ can be further stacked into
\begin{align} \label{hfd}
\widetilde{\mathbf{h}}_{fd} = \boldsymbol{\Phi} \overline{\mathbf{d}} + \widetilde{\mathbf{w}}_{fd},
\end{align}
where $\widetilde{\mathbf{h}}_{fd}\in \mathbb{C}^{KN^2\times 1}$, $\overline{\mathbf{d}} \in \mathbb{C}^{PQN\times 1}$ and $\widetilde{\mathbf{w}}_{fd} \in \mathbb{C}^{KN^2\times 1}$ are stacked by $\widetilde{\mathbf{h}}_{fd}(n)$, $\overline{\mathbf{d}} (n)$ and $\widetilde{\mathbf{w}}_{fd} (n)$ ($n=0,\ldots,N-1$), respectively, {{\color{black} and  $\boldsymbol{\Phi} = \left(\mathbf{I}_N\otimes (\overline{\mathbf{A}}_\nu^* \otimes \overline{\mathbf{A}}_\tau) \right)$.}
Note that ({\ref{hfd}) is an SSR problem and can be solved using the classical VBI method. 
However, since the number of grids $P$ and $Q$  is typically large,  the dimension of the measurement matrix becomes very high, leading to excessive computational complexity.

Next, we develop a two-layer VBI method to prevent multiplicative expansion over multiple signal dimensions, thereby reducing complexity substantially.

\subsection{Two-Layer SBL Framework}

In this subsection, we develop a two-layer SBL framework to  estimate the sparse matrix $\overline{\mathbf{D}}(n)$ in (\ref{Hfd}) efficiently and unambiguously. As is typical,  we use a tensor $\mathcal{D} \in \mathbb{C}^{P \times Q \times N}$ to denote the 3D sparse matrix constructed based on  $\overline{\mathbf{D}}(n)\in \mathbb{C}^{P\times Q}$ for $n =  0 ,\ldots,N-1$. Then,  the delay, fractional Doppler and integer Doppler  can be obtained based on the positions of the non-zero elements in the 3D matrix $\mathcal{D}$, which  correspond to the row indices, column indices and the slice indices, respectively. 

Since the main challenge is the estimation of Doppler, we place the Doppler estimation in the first layer. Hence, (\ref{Hfd}) can be rewritten by taking the Hermitian transposition as follows
{\begin{align}\label{Hfd2}
\mathbf{Y}(n) =\overline{\mathbf{A}}_{\nu} {\mathbf{X}}(n)\overline{\mathbf{A}}_{\tau}^H  + \mathbf{W}(n),
\end{align} }
where $\mathbf{Y}(n) = \widetilde{\mathbf{H}}_{f d}^H(n) \in \mathbb{C}^{K\times N}$,  ${\mathbf{X}}(n) = \overline{\mathbf{D}}(n)^H  \in \mathbb{C}^{Q\times P}$, $\overline{\mathbf{A}}_{\nu}\in \mathbb{C}^{K\times Q}$, $\overline{\mathbf{A}}_{\tau} \in \mathbb{C}^{N\times P}$ and $\mathbf{W}(n) = \widetilde{\mathbf{W}}_{f d}(n)^H\in \mathbb{C}^{K\times N}$. Note that the non-zero elements in $\mathbf{X}(n)$ is $h_l^*$ if and only if $n_lf_0 \approx \nu_l$ for $l = 1,\ldots,L$. 

Assume that the path coefficients $h_l^*(l = 1,\ldots,L)$, i.e. the  elements in $\mathbf{X}(n)$, follow the Gaussian distribution~\cite{Gaussian,Gaussian2}. Hence, the distribution of ${\mathbf{X}}(n)$ conditional on $\boldsymbol{\Gamma}^x(n)$ can be given by~\cite{bayesbase}
\begin{align} \label{Xn}
p({\mathbf{X}}(n) \mid \boldsymbol{\Gamma}^x(n)) & =\prod_{q=1}^{Q}\prod_{p=1}^{P} p\left({{X}}_{(q,p)}(n) \mid \gamma^x_{(q,p)}(n)\right) \notag \\ &=\prod_{q=1}^{Q}\prod_{p=1}^{P} \mathcal{C N}\left({{X}}_{(q,p)}(n) ; 0, \gamma^x_{(q,p)}(n)^{-1}\right),
\end{align}
where ${{X}}_{(q,p)}(n)$ is the $(q,p)$-the element in ${\mathbf{X}}(n)$, $\gamma^x_n(q,p)$ is the $(q,p)$-th element of $\boldsymbol{\Gamma}^x(n)$ and is the inverse of the variance of ${{X}}_{(q,p)}(n)$, which is called the precision in the SBL framework. The precision matrix $\boldsymbol{\Gamma}^x(n)$ follows the Gamma distribution, i.e.
\begin{align} \label{Gax}
p(\boldsymbol{\Gamma}^x(n))=\prod_{p = 1}^{P}\operatorname{Gamma}(\boldsymbol{\gamma}_p^x ; a, b),
\end{align}
where $\boldsymbol{\gamma}_p^x$ is the $p$-th column of $\boldsymbol{\Gamma}^x(n)$, $a$ is the shape parameter and $b$ is the inverse scale paramter in Gamma distribution. Since Gamma distribution is the conjugete {\it a priori} of the Gaussian distribution, the calculation of the {\it a posteriori} distribution can be simplified. Moreover,  the marginal distribution of $p({{X}}_{(q,p)}(n)) = \int p\left(\gamma^x_{(q,p)}(n)\right) p\left({{X}}_{(q,p)}(n) \mid \gamma^x_{(q,p)}(n)\right) d \gamma^x_{(q,p)}(n)$ is calculated as Student-t distribution by assuming that ${{X}}_{(q,p)}(n)$ obeys Gaussian distribution and $\gamma^x_{(q,p)}(n)$ obeys Gamma distribution, which ensures the sparsity of $\mathbf{X}(n)$~\cite{bayesbase2}.
 
{{\color{black}Then we assume that the noise matrix ${\mathbf{W}}(n)$ follows the Gaussian distribution with zero-mean and variance $\alpha^{-1}$, given by
\begin{align} \label{Wn}
p\left(\mathbf{W}(n)\right)=\prod_{m=0}^{M-1} \mathcal{C N}\left(0, \alpha^{-1} \mathbf{I}\right),
\end{align}
where $\alpha$ is the precision of the noise. Note that elements in ${\mathbf{W}}(n)$ are re-aligned by $\widetilde{w}_{f d}(n, m, k)$ with variance $\sigma^2$, so we have $\alpha = \sigma^{-2}$.} Generally, $\alpha$ is assumed to be Gamma distribution of
\begin{align}
p(\alpha)=\operatorname{Gamma}(\alpha ; a,b),
\end{align}
where $a$ is the shape parameter and $b$ is the inverse scale parameter in Gamma distribution.

However, since the matrix $\mathbf{X}(n)$ is multiplied by measurement matrices on both sides, we cannot directly provide the representation of $p\left(\mathbf{Y}(n) \mid  {\mathbf{X}}(n), \alpha\right)$, which is necessary for the estimation of $\mathbf{X}(n)$. Therefore, we propose a two-layer VBI method in order to estimate the sparse matrix $\mathbf{X}(n)$ without vecterizing $\mathbf{Y}(n)$. 

{\it 1) The first layer:}
The SSR problem in the first layer is as follows. Introducing $\mathbf{C}(n)  = {\mathbf{X}}(n)\overline{\mathbf{A}}_{\tau}^H \in \mathbb{C}^{Q\times N}$, (\ref{Hfd2}) can be written as
{\begin{align}\label{Hfd3}
\mathbf{Y}(n) =\overline{\mathbf{A}}_{\nu} {\mathbf{C}}(n)  + \mathbf{W}(n).
\end{align} }
Since $\mathbf{W}(n)$ follows the Gaussian distribution in (\ref{Wn}), we have
{\begin{align}
p\left(\mathbf{Y}(n) \mid  {\mathbf{C}}(n), \alpha\right)= \prod_{m=0}^{M-1}\mathcal{C N}\left(\overline{\mathbf{A}}_{\nu} \mathbf{c}_m(n), \alpha^{-1} \mathbf{I}\right) ,
\end{align}}
where $\mathbf{c}_m(n)$ is the $m$-th colomn of $\mathbf{C}(n)$. Next, we will provide the {\it a priori} distribution of $\mathbf{C}(n)$.
Since elements in $\mathbf{C}(n)$ are linear combinations of i.i.d. complex Gaussian variables in $\mathbf{X}(n)$, they obey the zero-mean Gaussian distribution, where the precision is based on $\boldsymbol{\Gamma}^x(n)$. Without loss of generality, we assume that the precision matrix of ${\mathbf{C}}(n)$ is  $\boldsymbol{\Gamma}^c(n)$. Hence, we have
\begin{align} \label{Cn}
p({\mathbf{C}}(n) \mid \boldsymbol{\Gamma}^c(n)) & =\prod_{m=1}^{M} \prod_{q=1}^{Q} p\left({{C}}_{(q,m)}(n) \mid \gamma^c_{(q,m)}(n)\right) \notag \\ &=\prod_{m=1}^{M} \prod_{q=1}^{Q} \mathcal{C N}\left({{C}}_{(q,m)}(n) ; 0, \gamma^c_{(q,m)}(n)^{-1}\right),
\end{align}
where  $\gamma^c_{(q,m)}(n)$ is the $(q,m)$-th element of $\boldsymbol{\Gamma}^c(n)$. Based on  the relation of {$\mathbf{C}(n)  = {\mathbf{X}}(n)\overline{\mathbf{A}}_{\tau}^H$}, we have 
{\begin{align} \label{Gacx}
\gamma^c_{(q,m)}(n) = (\sum_{p=1}^{P}\gamma^x_{(q,p)}(n)^{-1}a_{\tau}(m,p)a_{\tau}(m,p)^*)^{-1},
\end{align}}
{where $a_{\tau}(m,p)$ is the $(m,p)$-th element in $\overline{\mathbf{A}}_{\tau}$.} {{\color{black} Note that the calculation of $\gamma^c_{(q,m)}(n)$ is based on $\gamma^x_{(q,p)}(n)$'s rather than the conventional hyperprior in SBL framework and $\mathbf{X}(n)$ needs to be estimated from $\mathbf{C}(n)$.} Therefore, we introduce the second layer of VBI method. 

{\it 2) The second layer: }In the second layer of our proposed algorithm, we adopt VBI method on $\mathbf{C}(n)^H$ to further estimate $\mathbf{X}(n)$ and the precision matrix $\boldsymbol{\Gamma}^x(n)$. Specifically, we firstly provide the SSR problem in the second layer to be solved, which is given by
{{\color{black} \begin{align} \label{C2}
 \mathbf{C}(n)^H = \overline{\mathbf{A}}_{\tau} \overline{\mathbf{D}}(n) +\mathbf{E}(n),
\end{align}}
where {$\mathbf{C}(n)^H \in \mathbb{C}^{N\times Q}$, $\overline{\mathbf{D}}(n) \in \mathbb{C}^{P\times Q}$} and
$\mathbf{E}(n)\in \mathbb{C}^{N\times Q}$ is the estimation error. According to (\ref{Xn}) and (\ref{Gax}), we have the {\it a priori} distribution of {$\overline{\mathbf{D}}(n)$ and $\mathbf{\Gamma}^d(n)$} as
{\begin{align} \label{X2n}
p(\overline{\mathbf{D}}(n) \mid \boldsymbol{\Gamma}^d(n)) & \!=\!\prod_{q=1}^{Q}\prod_{p=1}^{P} p\left({\overline{D}}_{(p,q)}(n) \mid \gamma^{{d}}_{(p,q)}(n)\right) \notag \\ &\!=\! \prod_{q=1}^{Q}\prod_{p=1}^{P} \mathcal{C N}\! \left(\overline{D}_{(p,q)}(n) ; 0, \gamma^{{d}}_{(p,q)}(n)^{-1}\! \right)\!,
\end{align}}
and 
{\begin{align} \label{Gax2}
p(\boldsymbol{\Gamma}^d(n))=\prod_{q = 1}^{Q}\operatorname{Gamma}(\boldsymbol{\gamma}_{q}^d ; c, d),
\end{align}}
where {$\boldsymbol{\Gamma}^d(n)$ is the precision matrix of $\overline{\mathbf{D}}(n)$} and $c$ and $d$ are the shape parameter and inverse scale parameter of Gamma distribution, respectively. {Since ${\mathbf{X}}(n) = \overline{\mathbf{D}}(n)^H$, we have $\gamma^x_{(q,p)}(n) = \gamma^d_{(p,q)}(n)$.}

In the SBL framework, we can assume that $\mathbf{E}(n)$ obeys the complex Gaussian distribution of~\cite{bayesbase}
\begin{align} \label{En}
p(\mathbf{E}(n)) = \mathcal{CN}(0,\beta^{-1}\mathbf{I}),
\end{align} 
where $\beta$ is the precision of the estimation error and is assumed to obey the Gamma distribution of 
\begin{align}
p(\beta)=\operatorname{Gamma}(\beta ; e, f),
\end{align}
where $e$ and $f$ are the shape parameter and the inverse scale parameter of Gamma distribution, respectively.
According to (\ref{C2}) and (\ref{En}), we have the {\it a prior} distribution of {$\mathbf{C}(n)^H$} as
{\begin{align}
p\left(\mathbf{C}(n)^H \mid \overline{\mathbf{D}}(n), \beta\right)=\prod_{q=1}^{Q}\mathcal{C N}\left(\overline{\mathbf{A}}_{\tau} \overline{\mathbf{d}}_{q}(n), \beta^{-1} \mathbf{I}\right),
\end{align}
where $\overline{\mathbf{d}}_{q}(n)$ is the $q$-th {column} of $\overline{\mathbf{D}}(n)$.}

\subsection{Variational Bayesian Inference Principles}

\begin{figure}[t]
	\center{\includegraphics[width=9cm]{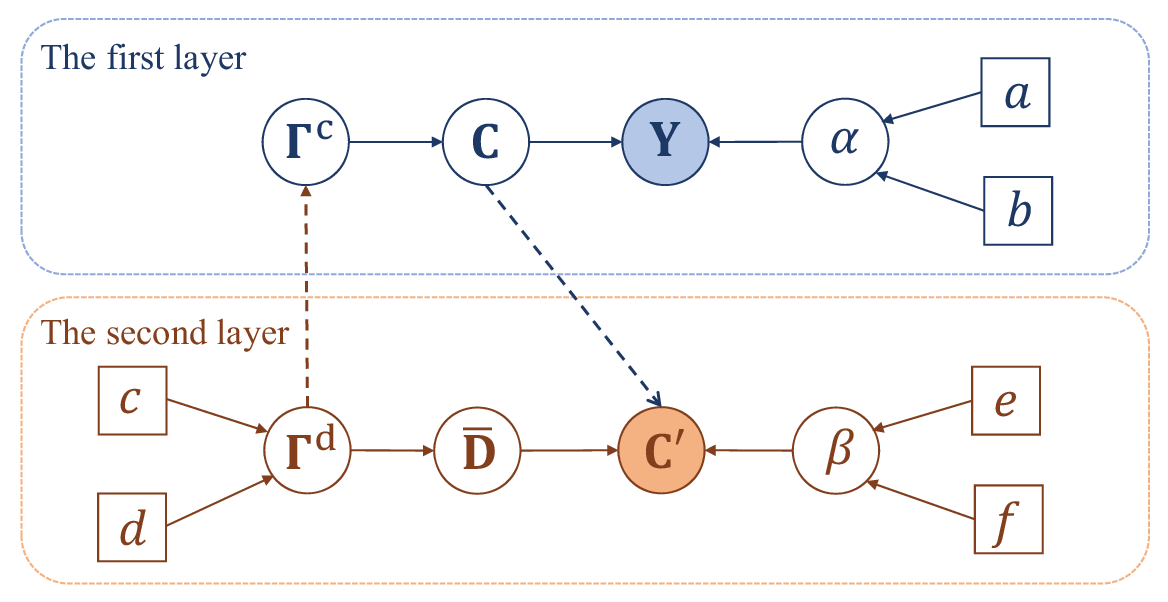}}
	\caption{The factor graph of the proposed two-layer VBI method.}
	\label{VBI1}
\end{figure}

The factor graph of our proposed two-layer VBI method is shown in Fig. \ref{VBI1}. The top diagram denotes the first layer, while the lower part denotes the second layer. The squares, the circles and the shaded circles represent constant values, hidden variables and observed values. For convenience, we use $\boldsymbol{\Xi} = \{\boldsymbol{\Xi}(n), n=0,\ldots,N-1 \}$ to denote the set of $N$ matrices $\boldsymbol{\Xi}(n)$, where $\boldsymbol{\Xi}$ can represent $\boldsymbol{\Gamma}^c$, $\mathbf{C}$, $\boldsymbol{\Gamma}^d$ , $\overline{\mathbf{D}}$ or $\mathbf{Y}$. The hidden variables to be estimated in the first layer of our proposed two-layer VBI method are denoted as $\boldsymbol{\Theta}_1 =  \{ \alpha, \boldsymbol{\Gamma}^c,\mathbf{C} \}$. The estimated mean of $\mathbf{C}(n)$, taken as a conjugate, serves as the observed variable in the second layer. 
Hence, the hidden variables to be estimated in the second layer of our two-layer VBI method are expressed as $\boldsymbol{\Theta}_2 = \{\beta,\boldsymbol{\Gamma}^d,\overline{\mathbf{D}} \}$.

 Based on the hierarchical model, the joint distribution of hidden variables for a given $n$ in the first layer is given by
\begin{align}
&p\left(\mathbf{Y}(n), \boldsymbol{\Theta}_1(n)\right) \notag \\ & =p\left(\mathbf{Y}(n) \mid \mathbf{C}(n), \alpha\right) p(\mathbf{C}(n) \mid \boldsymbol{\Gamma}^c(n)) p(\boldsymbol{\Gamma}^c(n)) p(\alpha).
\end{align}

 According to the mean field theory~\cite{bayesbase,bayesbase2,bayestutorial}, the joint distribution of all hidden variables is given by
\begin{align}
p\left(\mathbf{Y}, \boldsymbol{\Theta}_1\right) = \prod_{n=1}^{N} p\left(\mathbf{Y}(n), \boldsymbol{\Theta}_1(n)\right).
\end{align}
Similarly, we have the joint distribution of all hidden variables in the second layer as
{\begin{align}
p\left(\mathbf{C}', \boldsymbol{\Theta}_2\right) = \prod_{n=1}^{N} p\left(\mathbf{C}^H(n), \boldsymbol{\Theta}_2(n)\right), 
\end{align}}
where {$p\left(\mathbf{C}^H(n), \boldsymbol{\Theta}_2(n)\right)$} is the joint distribution of hidden variables corresponding to the Doppler index $n$ and can be written as
{\begin{align}
& p\left(\mathbf{C}^H(n), \boldsymbol{\Theta}_2(n)\right) \notag \\ &= p\left(\mathbf{C}^H(n) \mid \overline{\mathbf{D}}(n), \beta\right) p(\mathbf{C}^H(n) \mid \boldsymbol{\Gamma}^d(n)) p(\boldsymbol{\Gamma}^d(n)) p(\beta).
\end{align}}

In order to obtain the estimation of the delay and the Doppler, we need to calculate the maximum {\it a posteriori} (MAP) estimator of $p\left(\boldsymbol{\Theta}_1 \mid \mathbf{Y}\right)$ and {$p\left(\boldsymbol{\Theta}_2 \mid \mathbf{C}'\right)$}. However, this calculation relies on high-dimentional integration, which has excessive computational complexity. {{\color{black} Hence, we resort to the VBI method~\cite{bayestutorial,WXY,bayesbase,bayesbase2,Bayes} to iteratively approximate the {\it a posteriori} distributions of hidden variables $q(\boldsymbol{\Theta}_1)$ and  $q(\boldsymbol{\Theta}_2)$ instead of calculating $p\left(\boldsymbol{\Theta}_1 \mid \mathbf{Y}\right)$ and $p\left(\boldsymbol{\Theta}_2 \mid \mathbf{C}'\right)$ directly.}
With the proof established in Appendix \ref{app0}, the following proposition is provided for the approximations.
\begin{proposition} \label{P1}
	The stable solution of  $q(\boldsymbol{\Theta}_2)$ in the second layer can be obtained by alternately updating the following probability functions:
	{\begin{align}\label{beta2}
	q^{(t+1)}(\beta) \propto \exp \left(\langle\ln p(\mathbf{C}', \boldsymbol{\Theta}_2)\rangle_{q^{(t)}(\overline{\mathbf{D}}) q^{(t)}(\boldsymbol{\Gamma}^d)}\right),
	\end{align}
	\begin{align}\label{X2}
	q^{(t+1)}(\overline{\mathbf{D}}) \propto \exp \left(\langle\ln p(\mathbf{C}', \boldsymbol{\Theta}_2)\rangle_{q^{(t+1)}(\beta) q^{(t)}(\boldsymbol{\Gamma}^d)}\right),
	\end{align}
	\begin{align} \label{Gamma2}
	q^{(t+1)}(\boldsymbol{\Gamma}^d) \propto \exp \left(\langle\ln p(\mathbf{C}', \boldsymbol{\Theta}_2)\rangle_{q^{(t+1)}(\beta) q^{(t+1)}(\overline{\mathbf{D}})}\right),
	\end{align}}
	where $q^{(t)}(\cdot)$ denotes the probability function of variables in the $t$-th iteration and $\left\langle\ln p\left(\mathbf{C}', \boldsymbol{\Theta}_2\right)\right\rangle_{ q\left(x\right)q\left(y\right)}$ denotes the expection with respect to $x$ and $y$ {$(x,y = \overline{\mathbf{D}},\boldsymbol{\Gamma}^d,\beta)$}.

\end{proposition}

Similarly, we can provide the following proposition for approximation $q(\boldsymbol{\Theta}_1)$ as follows.
\begin{proposition} \label{P2}
	The stable solution of $q(\boldsymbol{\Theta}_1)$ in the first layer can be obtained by alternately updating the following probability functions:
	\begin{align} \label{alpha}
	q^{(i+1)}(\alpha) \propto \exp \left(\langle\ln p(\mathbf{Y}, \boldsymbol{\Theta}_1)\rangle_{q^{(i)}(\mathbf{C}) q^{(i)}(\boldsymbol{\Gamma}^c)}\right),
	\end{align}
	\begin{align} \label{C2n}
	q^{(i+1)}(\mathbf{C}) \propto \exp \left(\langle\ln p(\mathbf{Y}, \boldsymbol{\Theta}_1)\rangle_{q^{(i+1)}(\alpha) q^{(i)}(\boldsymbol{\Gamma}^c)}\right),
	\end{align}
	where we use $q^{(i)}(\cdot)$ to denote the probability function in the $i$-th iteration in the first-layer.
	
Note that $q(\boldsymbol{\Gamma}^c)$  cannot be directly obtained based on conventional VBI method. We have to calculate it based on {$\boldsymbol{\Gamma}^d(n)$}  obtained in the second layer as in (\ref{Gacx}), which indicates our proposed two-layer VBI method necessary.

\end{proposition}

In the next subsection, we will provide a detailed computation process for the two-layer VBI method by solving (\ref{beta2})-(\ref{C2n}).

\subsection{Update Details for SBL}
{Proposition \ref{P1} and \ref{P2} show that the approximations of $p\left(\boldsymbol{\Theta}_1 \mid \mathbf{Y}\right)$ and $p\left(\boldsymbol{\Theta}_2 \mid \mathbf{C}'\right)$ rely on the calculations of (\ref{beta2})-(\ref{C2n}). In this subsection, we will provide the recursive calculation of related variables to facilitate the approximations. }
We firstly provide the detailed process for (\ref{alpha})-(\ref{C2n}) in the first layer of our proposed algorithm.

\subsubsection{Update of $q(\alpha)$}
By solving (\ref{alpha}), the logarithm of $ q^{(i+1)}(\alpha)$ can be updated, as presented in Lemma 1.

\textit{Lemma 1:}
$\alpha$ follows the Gamma distribution of
\begin{align}
q^{(i+1)}(\alpha)=\operatorname{Gamma}\left(\alpha \mid \alpha_{\alpha}^{(i+1)}, b_{\alpha}^{(i+1)}\right),
\end{align}
where $a^{(i+1)}_{\alpha} = a+MNQ$ and {$b^{(i+1)}_{\alpha} = b+\sum_{n=1}^N\sum_{m=1}^N\left\|\mathbf{y}_m(n)-\overline{\mathbf{A}}_{\nu} \mathbf{u}_{\mathbf{c}_m(n)}^{(i)}\right\|_{2}^{2}+\operatorname{tr}\left(\overline{\mathbf{A}}_{\nu} \boldsymbol{\Sigma}_{\mathbf{c}_m(n)}^{(i)}\left(\overline{\mathbf{A}}_{\nu}\right)^{H}\right)$}, with $\mathbf{u}_{\mathbf{c}_m(n)}^{(i)}$ and $\boldsymbol{\Sigma}_{\mathbf{c}_m(n)}^{(i)}$ being the mean and the correlation matrix of $\mathbf{c}_m(n)$ at the $i$-th iteration. Their closed-form expressions will be given in the next lemma.

\textit{Proof:} See Appendix \ref{appA}.

Note that the mean of $\alpha$ can be calculated by
\begin{align} \label{alpha2}
\hat{\alpha}^{(i+1)}=\langle\alpha\rangle_{q^{(i+1)}(\alpha)}=\frac{a_{\alpha}^{(i+1)}}{b_{\alpha}^{(i+1)}},
\end{align}
which will be used in the updation of $q(\mathbf{C})$.

\subsubsection{Update of $q(\mathbf{C})$}
Similarly, by solving (\ref{C2n}), the logarithm of $ q^{(i+1)}(\mathbf{C})$ can be updated, as shown in Lemma 2.

\textit{Lemma 2:}
The $m$-th column of $\mathbf{C}$, which is denoted as
$\mathbf{c}_m(n)$, follows the Gaussian distribution of 
\begin{align}
q^{(i+1)}(\mathbf{c}_m(n))=\mathcal{C N}\left(\mathbf{c}_m(n) \mid \mathbf{u}_{\mathbf{c}_m(n)}^{(i+1)}, \boldsymbol{\Sigma}_{\mathbf{c}_m(n)}^{(i+1)}\right),
\end{align}
where
{\begin{align} \label{Sigmac}
\boldsymbol{\Sigma}_{\mathbf{c}_m(n)}^{(i+1)}=\left(\hat{\alpha}^{(i+1)} \overline{\mathbf{A}}_{\nu}^{H} \overline{\mathbf{A}}_{\nu}+\operatorname{diag}\{ (\hat{\boldsymbol{\gamma}}^c_m(n))^{(i)} \} \right)^{-1}
\end{align}}
and 
{\begin{align} \label{uc}
\mathbf{u}_{\mathbf{c}_m(n)}^{(i+1)}=\hat{\alpha}^{(i+1)} \boldsymbol{\Sigma}_{\mathbf{c}_m(n)}^{(i+1)} \overline{\mathbf{A}}_{\nu}^{H} \mathbf{y}_m(n)
\end{align}}
with $(\hat{\boldsymbol{\gamma}}^c_m(n))^{(i)}$ being the mean of ${\boldsymbol{\gamma}}^c_m(n)$ at the $i$-th iteration. Its closed-form expression will be given later.

\textit{Proof:} See Appendix \ref{appB}.

Hence, we have the estimation of $\mathbf{C}(n)$ as $\hat{\mathbf{C}}^{(i+1)}(n) = \left[\mathbf{u}_{\mathbf{c}_1(n)}^{(i+1)},\ldots,\mathbf{u}_{\mathbf{c}_M(n)}^{(i+1)} \right]$.

\subsubsection{Update of $q(\boldsymbol{\Gamma}^c(n))$ -- The Second-Layer VBI}

In our system model, $\boldsymbol{\Gamma}^c(n)$ is the precision matrix of $\mathbf{C}(n)$, which is multiplied by $\mathbf{X}(n)$ and the dictionary matrix $\overline{\mathbf{A}}_{\tau}^H$. Furthermore,  elements in $\mathbf{X}(n)$ are assumed to obey Gaussian distribution and its precision matrix $\boldsymbol{\Gamma}^x(n)$ obeys Gamma distribution, which excites the sparsity of $\boldsymbol{\Gamma}^x(n)$~\cite{bayesbase2}. Hence, we cannot assume that $\boldsymbol{\Gamma}^c(n)$ follows a specific distribution. Instead, we obtain the mean of $\boldsymbol{\Gamma}^c(n)$ from $\boldsymbol{\Gamma}^x(n)$ based on (\ref{Gacx}), where we need to introduce the second layer of VBI method.

Specifically, we have $\hat{\mathbf{C}}^{(i+1)}(n)^H$ as the observed matrix in (\ref{C2}). Therefore, we can adopt the VBI method on (\ref{C2}) to obtain the matrices $\mathbf{X}(n)$ and $\boldsymbol{\Gamma}^x(n)$. Similarly, we need to update the statistic information of $\beta$, $\overline{\mathbf{D}}(n)$ and $\boldsymbol{\Gamma}^d(n)$.

\textit{Lemma 3:} The mean of $\beta$  can be expressed as
\begin{align} \label{beta}
\hat{\beta}^{(t+1)}=\langle\beta\rangle_{q^{(t+1)}(\beta)}=\frac{a_{\beta}^{(t+1)}}{b_{\beta}^{(t+1)}},
\end{align}
where $a^{(t+1)}_{\beta} = a+NPQ$ and $b^{(t+1)}_{\beta} = b+\sum_{n=1}^N\sum_{q=1}^Q\left\|\overline{\mathbf{d}}_{q}(n)-\overline{\mathbf{A}}_{\tau} \mathbf{u}_{\overline{\mathbf{d}}_{q}(n)}^{(t)}\right\|_{2}^{2}+\operatorname{tr}\left(\overline{\mathbf{A}}_{\tau} \boldsymbol{\Sigma}_{\overline{\mathbf{d}}_{q}(n)}^{(t)}\left(\overline{\mathbf{A}}_{\tau}\right)^{H}\right)$, with $\mathbf{u}_{\overline{\mathbf{d}}_{q}(n)}^{(t)}$ and $\boldsymbol{\Sigma}_{\overline{\mathbf{d}}_{q}(n)}^{(t)}$ being the mean and the correlation matrix of the $q$-th column of $\overline{\mathbf{D}}(n)$, which is denoted as $\overline{\mathbf{d}}_{q}(n)$, at the $t$-th iteration, respectively. Their closed-form expressions will be given in the next lemma.

\textit{Proof:} See Appendix \ref{appC}.

\textit{Lemma 4:} The mean and the correlation matrix of $\overline{\mathbf{d}}_{q}(n)$ can be expressed as
\begin{align} \label{ux}
\mathbf{u}_{\overline{\mathbf{d}}_{q}(n)}^{(t+1)}=\hat{\beta}^{(t+1)} \boldsymbol{\Sigma}_{\overline{\mathbf{d}}_{q}(n)}^{(t+1)} \overline{\mathbf{A}}_{\tau}^{H} \mathbf{c}_{q}^H(n),
\end{align}
and
\begin{align} \label{Sigmax}
\boldsymbol{\Sigma}_{\overline{\mathbf{d}}_{q}(n)}^{(t+1)}=\left(\hat{\beta}^{(t+1)} \overline{\mathbf{A}}_{\tau}^{H} \overline{\mathbf{A}}_{\tau}+\operatorname{diag}\{ (\hat{\boldsymbol{\gamma}}^d_{q}(n))^{(t)} \} \right)^{-1},
\end{align}
with {$(\hat{\boldsymbol{\gamma}}^d_{q}(n))^{(t)}$} being the estimation of the $q$-th column of {${\boldsymbol{\Gamma}}^d(n)$} at the $t$-th iteration and {$\mathbf{c}_{q}^H(n)$ being the $q$-th row of matrix $\mathbf{C}$}.

\textit{Proof:} See Appendix \ref{appD}.

Hence, we have the estimation of $\overline{\mathbf{D}}(n)$ as $\hat{{\mathbf{D}}}^{(t+1)}(n) = \left[{\overline{\mathbf{d}}_{1}(n)}^{(t+1)},\ldots,\overline{\mathbf{d}}_{Q}(n)^{(t+1)} \right]$.

\textit{Lemma 5:} The $(p,q)$-th element of $\boldsymbol{\Gamma}^d$, which is denoted as {$\gamma^d_{(p,q)}(n)$}, follows the Gamma distribution of
\begin{align}
q^{(t+1)}\left(\gamma^d_{(p,q)}(n)\right)=\operatorname{Gamma}\left(\gamma^d_{(p,q)}(n)  \mid a_{\gamma^d_{(p,q)}(n)}^{(t+1)}, b_{\gamma^d_{(p,q)}(n)}^{(t+1)} \right).
\end{align}
Hence, the mean of ${\gamma}^d_{(p,q)}(n)$ is 
\begin{align} \label{gamma}
{\hat{\gamma}^{d}_{(p,q)}}(n)^{(t+1)}=\left\langle{\gamma}^d_{(p,q)}(n)\right\rangle_{q^{(t+1)}\left(\boldsymbol{\Gamma}^d\right)}=\frac{a_{\gamma^d_{(p,q)}(n)}^{(t+1)}}{b_{\gamma^d_{(p,q)}(n)}^{(t+1)}},
\end{align}
where $a_{\gamma^d_{(p,q)}(n)}^{(t+1)} = a+1$ and  $b_{\gamma^d_{(p,q)}(n)}^{(t+1)} = b+|\mathbf{u}_{\overline{\mathbf{d}}_{(p,q)}(n)}^{(t+1)}|^2+\boldsymbol{\Sigma}_{\overline{\mathbf{d}}_{(p,q)}(n)}^{(t+1)}$. $\mathbf{u}_{\overline{\mathbf{d}}_{(p,q)}(n)}^{(t+1)}$ denotes the $p$-th element of $\mathbf{u}_{\overline{\mathbf{d}}_{q}(n)}^{(t+1)}$ and $\boldsymbol{\Sigma}_{\overline{\mathbf{d}}_{(p,q)}(n)}^{(t+1)}$ denotes the $(p,p)$-th element of $\boldsymbol{\Sigma}_{\overline{\mathbf{d}}_{q}(n)}^{(t+1)}$.

\textit{Proof:} See Appendix \ref{appE}.

Then we can calculate the estimation of  $\gamma^c_{(q,m)}(n)$ based on (\ref{Gacx}) and obtain $\operatorname{diag}\{ (\hat{\boldsymbol{\gamma}}^c_m(n))^{(i)} \}$ in (\ref{Sigmac}).

\begin{algorithm}[!t]
	\caption{The proposed two-layer VBI method.}
	\renewcommand{\algorithmicrequire}{\textbf{Input:}}
	\renewcommand{\algorithmicensure}{\textbf{Initialization:}}

	\begin{algorithmic}[1]
			\REQUIRE{ Re-aligned channel matrix  $\widetilde{\mathbf{H}}_{f d}(n)$, noise matrix $\widetilde{\mathbf{W}}_{f d}(n)$ for $ n =0,\ldots,N-1$, measurement matrices $\overline{\mathbf{A}}_{\nu}$ and $\overline{\mathbf{A}}_{\tau}$. 	}
			\ENSURE{Initial values of  $\hat{\alpha}^{(0)}$,  $\mathbf{u}_{\mathbf{c}_m(n)}^{(0)}$, $\boldsymbol{\Sigma}_{\mathbf{c}_m(n)}^{(0)}$,  $\hat{\beta}^{(0)}\!$, $\mathbf{u}_{\overline{\mathbf{d}}_{q}(n)}^{(0)}$,  $\boldsymbol{\Sigma}_{\overline{\mathbf{d}}_{q}(n)}^{(0)}$, $\hat{\gamma}^d_{(p,q)}(n)^{(0)}$  and $\hat{\gamma}^c_{(q,m)}(n)^{(0)}$. The error threshold $\zeta_1 = 10^{-5}$ for the first layer and $\zeta_2 = 10^{-5}$ \\for the second layer, maximum iteration $N_{max1} = 167$ and $N_{max2}=167$, iteration counter $i=1$ for the first layer and $t=1$ for the second layer. }
			
			\REPEAT
			
			\STATE Update $\hat{\alpha}^{(i+1)}$ by (\ref{alpha2}).
			\STATE Update $\mathbf{u}_{\mathbf{c}_m(n)}^{(i+1)}$ and $\boldsymbol{\Sigma}_{\mathbf{c}_m(n)}^{(i+1)}$ by (\ref{uc}) and (\ref{Sigmac}), respectively.
			
			\REPEAT
			\STATE Update $\hat{\beta}^{(t+1)}$ by (\ref{beta}).
			\STATE Update $\mathbf{u}_{\mathbf{x}_{2,q}(n)}^{(t+1)}$ and $\boldsymbol{\Sigma}_{\mathbf{x}_{2,q}(n)}^{(t+1)}$ by (\ref{ux}) and (\ref{Sigmax}), respectively.
			\STATE Update $\hat{\gamma}^x_{2,(p,q)}(n)^{(t+1)}$ by (\ref{gamma}).
			\STATE Set $t = t+1$.
			\UNTIL $\sum_{n=1}^{N}\frac{\left\|\boldsymbol{\Gamma}_2^x(n)^{(t+1)}-\boldsymbol{\Gamma}_2^x(n)^{(t)}\right\|_2^2}{\left\|\boldsymbol{\Gamma}_2^x(n)^{(t)}\right\|_2^2} \leq \zeta_2$ or $t\geq N_{max2}$.

			\STATE Update $\gamma^c_{(q,m)}(n)^{(i+1)}$ by (\ref{Gacx}).
			\STATE Set $i=i+1$.

			\UNTIL $\sum_{n=1}^{N}\frac{\left\|\boldsymbol{\Gamma}^c(n)^{(i+1)}-\boldsymbol{\Gamma}^c(n)^{(i)}\right\|_2^2}{\left\|\boldsymbol{\Gamma}_2^x(n)^{(i)}\right\|_2^2} \leq \zeta_1$ or $i\geq N_{max1}$.

	\end{algorithmic}	
	
\KwOut{Sparse matrices $\mathbf{X}(n),\ n=0,\ldots,N-1 $}

\end{algorithm}}

\textbf{Algorithm 1} summarizes the proposed two-layer VBI estimation method.  We start with updating the variables  $\hat{\alpha}^{(i+1)}$, $\mathbf{u}_{\mathbf{c}_m(n)}^{(i+1)}$ and $\boldsymbol{\Sigma}_{\mathbf{c}_m(n)}^{(i+1)}$	in the first layer, which is referred to as the outer iteration including Steps $2$, $3$ and $10$. To update the precision $\gamma^c_{(q,m)}(n)^{(i+1)}$, we resort to the second layer, known as the inner iteration covering Steps $5$-$8$.	In the second layer, we update $\hat{\beta}^{(t+1)}$, $\mathbf{u}_{\overline{\mathbf{d}}_{q}(n)}^{(t+1)}$,  $\boldsymbol{\Sigma}_{\overline{\mathbf{d}}_{q}(n)}^{(t+1)}$ and $\hat{\gamma}^d_{(p,q)}(n)^{(t+1)}$ based on (\ref{beta}), (\ref{ux}), (\ref{Sigmax}) and (\ref{gamma}) using the VBI method. After convergence or {{\color{black}when the iteration number reaches the maximum iteration number\footnote{{\color{black}The maximum parameter iteration count in the simulation is set to $500$. Since each iteration updates three parameters, the maximum iteration number in the first layer and the second layer are $N_{max1} = 500/3 \approx 167$ and $N_{max2}=167$, respectively.},} we can calculate $\gamma^c_{(q,m)}(n)^{(i+1)}$ based on the final $\hat{\gamma}^d_{(p,q)}(n)^{(t+1)}$. Once the outer iteration converges, we obtain the final estimation of $\mathbf{X}(n)$. The delay $\tau_l$, the fractional Doppler $\xi_l$ and the integer Doppler $n_l$ can be estimated from the row indices, column indices and slice indices of the non-zero elements in $\mathbf{X}(n)$. Furthermore, since these three variables indicate the location of a specific non-zero element, they are matched naturally.

So far, we have developed a two-layer VBI-based estimation framework for estimating sensing parameters in time-varying channels. {{\color{black}The developed method addresses the parameter mismatching issue in conventional methods by estimating all parameters together unambiguously and achieves better performance compared with the conventional sensing methods in~[31] because of the reasonable {\it a priori} distribution assumption.} It also reduces the computational complexity of SSR substantially by avoiding the high-dimensional signal stacking as in conventional SSR. Nevertheless, we note that the sensing parameters' value regions can be substantially narrowed down, combined with other sensing methods. This can further reduce the computational complexity of the proposed two-layer VBI sensing methods. Next, we provide a simplified two-stage VBI design.

\section{Simplified Two-Stage VBI} \label{S4}

In this section, we propose a simplified two-stage VBI algorithm. {{\color{black} By reorganizing the signals in {(\ref{Hfd})}, we show that targets' delays can be readily estimated using the conventional estimation method in Stage 1. One of the most popular estimation techniques is the MUSIC algorithm, which has also been widely used in ISAC parameter estimations lately~\cite{ISACBayes4,MUSIC2,MUSIC4,MUSIC5}.} Then, in Stage 2, we use MUSIC-estimated delays in the two-layer VBI framework to estimate the Doppler frequencies. 

In Stage 1, we firstly introduce a new tactic to reformulate $\widetilde{\mathbf{H}}_{f d}(n)$ for  $n=0,\ldots,N-1$ for the delay estimation. Specifically, we propose to construct a new matrix 
$\mathbf{H}_1=[\widetilde{\mathbf{H}}_{f d}^*(0), \ldots, \widetilde{\mathbf{H}}_{f d}^*(N-1)]^T\in \mathbb{C}^{KN\times N}$ by stacking $\widetilde{\mathbf{H}}_{f d}(n)^H$ for $n = 0,\ldots,N-1$.
Based on (\ref{Hfd}), $\mathbf{H}_1$ can be further rewritten as
\begin{align} \label{H2}
\mathbf{H}_1&=\left[\begin{array}{c} \widetilde{\mathbf{H}}_{f d}^H(0) \\ 
\vdots \\ 
\widetilde{\mathbf{H}}^H_{f d}(N-1)\end{array}\right] \\
&=\left[\begin{array}{c} \mathbf{A}_{\nu} {\mathbf{D}(0)}^H \mathbf{A}_{\tau}^{H} \\ 
\vdots \\ 
\mathbf{A}_{\nu} {\mathbf{D}(N-1)}^H \mathbf{A}_{\tau}^{H}\end{array}\right] + \left[\begin{array}{c} \widetilde{\mathbf{W}}_{f d}(0)^H \\ 
\vdots \\ 
\widetilde{\mathbf{W}}_{f d}(N-1)^H\end{array}\right] \\
&=\left[\begin{array}{ccc}\mathbf{A}_{\nu} &  & 
\\  & \ddots & 
\\ &  & \mathbf{A}_{\nu}\end{array}\right]
\left[\begin{array}{c} {\mathbf{D}(0)}^H \\ 
\vdots \\ 
{\mathbf{D}(N-1)}^H\end{array}\right]
\mathbf{A}_{\tau}^{H} + \mathbf{W}_1,
\end{align}
where $\mathbf{W}_1 = [\widetilde{\mathbf{W}}_{f d}^*(0), \ldots, \widetilde{\mathbf{W}}_{f d}^*(N-1)]^T\in \mathbb{C}^{KN\times N}$ is the stacked noise matrix.

The following correlation matrix $\mathbf{R}_1$ can be derived as
\begin{align} \label{Corr}
\mathbf{R}_1 & =\mathbb{E} \{{\mathbf{H}_1}^H\mathbf{H}_1\}  \notag \\
& =\mathbf{A}_{\tau}
\left(\sum_{n=0}^{N-1}\mathbf{D}(n)\mathbf{A}_{\nu}^H \mathbf{A}_{\nu} \mathbf{D}(n)^H \right) \mathbf{A}_{\tau}^H+ \sigma^2 \mathbf{I}_{M}.
\end{align}
We see that $\mathbf{R}_1$ can be seen as a correlation of the signal matrix underlain by the delay steering vectors. This enables us to use the MUSIC algorithm to estimate the delays conveniently. The eigenvalue decomposition (EVD) of $\mathbf{R}_1$ can be given by
\begin{align} \label{EVD}
\mathbf{R}_1 = \mathbf{U}_1\boldsymbol{\Sigma}\mathbf{U}_1^H,
\end{align}
where $\boldsymbol{\Sigma}\in \mathbb{R}^{N\times N}$ is a digonal matrix with eigenvalues arranged in the descending order and $\mathbf{U}_1\in \mathbb{C}^{N\times N}$ is a unitary matrix. The null-space of $\mathbf{U}_1$ can be constructed as $\mathbf{U}_N = \mathbf{U}_1(:,L+1:N)$, then the delays can be estimated by identifying the peaks in the following MUSIC spectrum
\begin{align} \label{MUS}
P_{\mathrm{MUSIC}}(\tau_p) = \frac{1}{\mathbf{a}^H(\tau_p) \mathbf{U}_N \mathbf{U}_N^H \mathbf{a}(\tau_p)},
\end{align}
{{\color{black}where  $\mathbf{a}(\tau_p) = \left[e^{-j 2 \pi 0 f_{0} \tau_{p}},\ldots, e^{-j 2 \pi (N-1) f_{0} \tau_{p}} \right]^T\in \mathbb{C}^{N\times 1}$ is the $p$-th column of $\overline{\mathbf{A}}_{\tau}$.} By taking the $L$ delays corresponding to the $L$ largest peak values, we can obtain the delay estimates $\tau_l$~\cite{MUSIC,MUSIC2}. This completes Stage 1 processing.

Proceeding to Stage 2, by plugging the delay estimation into the dictionary matrix, we can obtain the matrix $\hat{\mathbf{A}}_{\tau}\in \mathbb{C}^{N\times L}$, where the $(m,l)$-th element is expressed as $e^{-j 2 \pi m f_{0} \tau_{l}}$. This allows us to suppress the second layer of the proposed VBI process, simplifying the system model in (\ref{Hfd2}) as follows:
\begin{align}\label{Hfd4}
\mathbf{Y}(n) =\boldsymbol{\Phi}_1 {\mathbf{X}}'(n)\hat{\mathbf{A}}_{\tau}^H   + \mathbf{W}(n),
\end{align} 
where ${\mathbf{X}}'(n) \in \mathbb{C}^{Q\times L}$ is a sparse matrix to be estimated. Note that $\hat{\mathbf{A}}_{\tau}$ is a known matrix, unlike the original dictionary matrix in (\ref{Hfd2}).

Next, we right-multiply both sides of the equation by  $\hat{\mathbf{A}}_{\tau}^{\dagger} = \hat{\mathbf{A}}_{\tau} \left(\hat{\mathbf{A}}_{\tau}^H \hat{\mathbf{A}}_{\tau} \right)$, yielding:
\begin{align} \label{Y2}
\mathbf{Y}'(n) =\boldsymbol{\Phi}_1 \mathbf{X}'(n)   + \mathbf{W}'(n),
\end{align}
where $\mathbf{Y}'(n) = \mathbf{Y}(n)\hat{\mathbf{A}}_{\tau}^{\dagger}$ and $\mathbf{W}'(n) = \mathbf{W}(n)\hat{\mathbf{A}}_{\tau}^{\dagger}$. The second layer of the VBI framework is removed by the known delays, leaving only the first layer, which corresponds to the traditional VBI framework. By adopting VBI method to solve the MM-SSR problem in (\ref{Y2}), we can estimate the sparse matrix $\mathbf{X}'(n) , n = 0,\ldots,N-1$. The integer Dopplers and fractional Dopplers are then extracted from the slice indices and row indices of the non-zero elements in $\mathbf{X}'(n)$. Additionally, the non-zero elements in the $l$-th column correspond to the Doppler associated with the 
$l$-th delay in the multi-target scenario. To summarize  the simplified two-stage VBI method at a glance, we represent it in \textbf{Algorithm 2}. 


\begin{algorithm}[!t]
	\caption{The simplified two-stage VBI method.}
	
	\KwIn{ Re-aligned channel matrix  $\widetilde{\mathbf{H}}_{f d}(n)$ for $ n =0,\ldots,N-1$,  measurement matrices $\boldsymbol{\Phi}_1 = \overline{\mathbf{A}}_{\nu}$ and $\boldsymbol{\Phi}_2 = \overline{\mathbf{A}}_{\tau}$. 
	}

	{\begin{algorithmic}[1]
			
			\item[]\textbf{Stage 1: Delay Estimation}
			\STATE Stack $\widetilde{\mathbf{H}}_{f d}(n)$ for $ n =0,\ldots,N-1$ as $\mathbf{H}_1$.
			\STATE Calculate the correlation matrix $\mathbf{R}_1$ of $\mathbf{H}_1$ by (\ref{Corr}).
			\STATE Adopt MUSIC method to estimate the delays $\tau_l$ by (\ref{EVD}) and (\ref{MUS}).
			\item[]\textbf{Stage 2: Doppler Estimation}
			\STATE Obtain $\hat{\mathbf{A}}_{\tau}$ and then obtain (\ref{Y2}) by right-multiplying (\ref{Hfd4}) by $\hat{\mathbf{A}}_{\tau}^{\dagger}$.
			\STATE Adopt VBI method on (\ref{Y2}) and obtain the Doppler estimates $\nu_l$.

	\end{algorithmic}}	
	
	\KwOut{The delay $\tau_l$ and the Doppler $\nu_l$ for $l=1,\ldots,L$.}
	
\end{algorithm}

{{\color{black} \emph{Remark 2}:  As mentioned in \textit{Remark 1}, the Doppler frequency needs to be treated as integer and fractional parts while the delay does not have the issue. Therefore, we use MUSIC for delay estimation to reduce computational complexity. However, MUSIC cannot be used for Doppler estimation, as the two distinct Doppler components can lead to the challenging parameter association and mismatch issues in multiple-target scenarios.}

\section{Cramér-Rao Bound} \label{S5}

Performance analysis is performed in this section by deriving the CRB for the delay $\tau_{l}$'s, the Doppler $\nu_{l}$'s and the path coefficients $h_l$'s for  the fast-fading channel. We use $\boldsymbol{\theta} = \{\tau_1,\ldots,\tau_L,\nu_1,\ldots,\nu_L,\operatorname{Re}\{h_1\},\ldots,\operatorname{Re}\{h_L\},\operatorname{Im}\{h_1\},\ldots,\\ \operatorname{Im}\{h_L\} \}$ to denote the unknown patameters for convenience.

Based on the system model in Section \ref{S2}, the input-output relationship for the $k$-th OFDM block can be written as
\begin{align}
\mathbf{r}(k) = \mathbf{H}_{fd}(k)\mathbf{s}(k)+\mathbf{w}_1(k),
\end{align}
where $\mathbf{w}_1(k)$ is the noise vector added at the receiver following Gaussian distribution with zero mean and variance $\sigma_1^2$. After collecting $K$ observations, we have the received matrix as
\begin{align}
\mathbf{R} &= \left[\mathbf{r}(1),\ldots,\mathbf{r}(K)\right] \notag \\
& = \left[\mathbf{H}_{fd}(1)\mathbf{s}(1),\ldots,\mathbf{H}_{fd}(K)\mathbf{s}(K)\right]+\left[\mathbf{w}_1(1),\ldots,\mathbf{w}_1(K)\right].
\end{align}

{Similarly, there is also sensing noise in the sensing parameter estimation process as shown in (\ref{Hfd2}). With the use of specialized pilot in the OFDM block~\cite{ZHY2}, the sensing noise can be assumed to follow a Gaussian distribution with zero mean and variance $\sigma_2^2$. By setting the mean power of the pilots $\mathbf{s}(k)$ as unity, the  total noise variance is given by $\sigma^2 = \sigma_1^2 + \sigma_2^2$.}

The log-likelihood function of $\mathbf{R}$ w.r.t. $\boldsymbol{\theta}$ is given by 
{\begin{align} \label{F}
\mathcal{F} &= \operatorname{ln}p(\mathbf{R};\boldsymbol{\theta}) \notag \\
& \propto KN\operatorname{ln}\sigma^2 - \sum_{k=1}^{K}\{\frac{1}{\sigma^2} \|\mathbf{r}(k) - \mathbf{H}_{fd}(k)\mathbf{s}(k)\|_2^2 \}.
\end{align}}
Based on (\ref{F}), we can obtain the first order derivatives w.r.t. $\theta_i$ for $i=1,\ldots,4L$ as
\begin{align}
\frac{\partial \mathcal{F}}{\partial \theta_i}=-\frac{1}{\sigma^2}\sum_{k=1}^{K}(\mathbf{r}(k) - \mathbf{H}_{fd}(k)\mathbf{s}(k))^{{H}}  \frac{\partial \mathbf{H}_{fd}(k)}{\partial \theta_i} \mathbf{s}(k)\notag \\ -\frac{1}{\sigma^2} \left(\frac{\partial \mathbf{H}_{fd}(k)}{\partial \theta_i} \mathbf{s}(k)\right)^{{H}} (\mathbf{r}(k) - \mathbf{H}_{fd}(k)\mathbf{s}(k)).
\end{align}
We can easily obtain that $\mathbb{E}\{\frac{\partial \mathcal{F}}{\partial \theta_i} \} = 0$ for $\forall i$ since $\mathbb{E}\{\mathbf{r}(k) - \mathbf{H}_{fd}(k)\mathbf{s}(k) \} = 0$. Therefore, the second order partial derivatives of $\mathcal{F}$ w.r.t. $\theta_i$ can be calculated as
\begin{align}
&\mathbb{E}\left\{\frac{\partial^2 \mathcal{F}}{\partial \theta_i \partial \theta_j^*}\right\} \notag \\ &=-\frac{2}{\sigma^2}\sum_{k=1}^{K}  \operatorname{Re}\left\{\left(\frac{\partial \mathbf{H}_{fd}(k)}{\partial \theta_j} \mathbf{s}(k)\right)^{\mathrm{H}}  \frac{\partial \mathbf{H}_{fd}(k)}{\partial \theta_i} \mathbf{s}(k)\right\}.
\end{align}

For convenience, we define that $\mathbf{T}_l(k) = \frac{\partial \mathbf{H}_{fd}(k)}{\partial \tau_l} $, $\mathbf{V}_l(k) = \frac{\partial \mathbf{H}_{fd}(k)}{\partial \nu_l} $, $\mathbf{R}_l(k) = \frac{\partial \mathbf{H}_{fd}(k)}{\partial \operatorname{Re}\{h_{l'} \}}$ and $\mathbf{I}_l(k) = \frac{\partial \mathbf{H}_{fd}(k)}{\partial \operatorname{Im}\{h_{l'} \}}$.
Next, we calculate $\mathbf{T}_l(k)$ as an example. With the same logic, $\mathbf{V}_l(k)$, $\mathbf{R}_l(k)$ and $\mathbf{I}_l(k)$ can be obtained, which will not be detailed for conciseness. Based on (\ref{Hfdnm}), $\mathbf{T}_l(k)$ can be expanded as
\begin{align}
\mathbf{T}_l(k) = \frac{\partial \mathbf{H}_{fd}(k)}{\partial \tau_l} = \left[\frac{\partial \mathbf{h}_{fd,1}(k)}{\partial \tau_l},\ldots,\frac{\partial \mathbf{h}_{fd,M}(k)}{\partial \tau_l} \right],
\end{align}
where $\frac{\partial \mathbf{h}_{fd,m}(k)}{\partial \tau_l}$ denotes the first order derivative of the $m$-th column of $\mathbf{H}_{fd}(k)$ w.r.t. $\tau_l$. However, we cannot directly differentiate $\mathbf{h}_{fd,m}(k)$ w.r.t. $\nu_l$ since we re-align the channel matrix $\mathbf{H}_{fd}$ for sensing and $\frac{\partial g_2(nf_0-\nu_l)}{\partial \nu_l}$ cannot be directly obtained. Therefore, we adopt IFFT and use the permutation matrix $\mathbf{P}^{(m-1)}$ mathematically to depict the realignment.
The $m$-th column of the re-aligned channel matrix in the frequency-time domain is given by
\begin{align}
\tilde{\mathbf{h}}_{ft,m}(k) = \mathbf{F}^H \mathbf{P}^{(m-1)}\mathbf{h}_{fd,m}(k),
\end{align}
where $\mathbf{P}^{(m-1)}$ is the permutation matrix that upwards the elements by $(m-1)$ positions.
Hence, we have
\begin{align}
\mathbf{h}_{fd,m}(k) =  {(\mathbf{P}^{(m-1)})}^{-1}\mathbf{F}\tilde{\mathbf{h}}_{ft,m}(k),
\end{align}
and the $m$-th column of $\mathbf{T}_l(k)$ 
\begin{align}
\frac{\partial \mathbf{h}_{fd,m}(k)}{\partial \tau_l} ={(\mathbf{P}^{(m-1)})}^{-1}\mathbf{F}  \frac{\partial \tilde{\mathbf{h}}_{ft,m}(k)}{\partial \tau_l},
\end{align}
where $\frac{\partial \tilde{\mathbf{h}}_{ft,m}(k)}{\partial \tau_l}$ is easy to calculate.

With $\mathbf{T}_l(k)$, $\mathbf{V}_l(k)$, $\mathbf{R}_l(k)$ and $\mathbf{I}_l(k)$ calculated,  the $(i,j)$-th element of the FIM $\mathbf{J}(\boldsymbol{\theta})$ can be calculated as~\cite{estimate}}
\begin{align}
J_{i,j} = -\mathbb{E}\left\{\frac{\partial^2 \mathcal{F}}{\partial \theta_i \partial \theta_j^*}\right\}, 1 \leq i, j \leq 4 L.
\end{align}
and the MSE of the unbiased estimator $\hat{\mathbf{\theta}}$ is lower-bounded by
\begin{align}
\mathbb{E}\left\{(\hat{\boldsymbol{\theta}}-\boldsymbol{\theta})(\hat{\boldsymbol{\theta}}-\boldsymbol{\theta})^{\mathrm{T}}\right\} \succeq \boldsymbol{J}^{-1}(\boldsymbol{\theta})
\end{align}
where the diagonal elements of $\boldsymbol{J}^{-1}(\boldsymbol{\theta})$ are the CRLBs for $\tau_l$, $\nu_l$, $\operatorname{Re}\{h_l \}$ and $\operatorname{Im}\{h_l \}$.

\section{Simulation Results}\label{S6}
	In this section, simulation results are provided to validate the proposed designs compared with the prior art.  Unless otherwise specified, simulation parameters are set as follows. The number of subcarriers is set as $N = 8$, while the number of OFDM blocks is set as $K=8$. {The carrier frequency is $f_c = 150$ GHz and the subcarrier spacing is $f_0 = 15$ kHz.} The number of targets is $L=3$ and the scattering coefficients are generated according to the complex Gaussian distribution of $\mathcal{CN}(0,1/L)$. The maximum delay and Doppler are set within $\left[0, 3T_0 \right]$ and $\left[-4f_0, 4f_0 \right]$, respectively. The mean squared error (MSE) for the Doppler and delay  is defined as
	\begin{align}
	\mathrm{MSE}_{{x}}=\frac{\sum_{l=1}^L \| \hat{x}_l - x_l\|^2}{L}, {x} = {\nu},{\tau},
	\end{align}
	where $\hat{x}_l$ is the estimation of $x_l$.
	The MSE below is calculated over  10000 independent realizations. 
	
	{{\color{black}As for the initial values of the proposed two-layer VBI method, simulation shows that our proposed method is robust to the initial values. Hence, we assume that $\hat{\alpha}^{(0)} = 1$, $\hat{\beta}^{(0)} = 1$,  $\mathbf{u}_{\mathbf{c}_m(n)}^{(0)}=\hat{\alpha}^{(i+1)} \boldsymbol{\Sigma}_{\mathbf{c}_m(n)}^{(0)} \overline{\mathbf{A}}_{\nu}^{H} \mathbf{y}_m(n)$,  $\boldsymbol{\Sigma}_{\mathbf{c}_m(n)}^{(0)}=\left(\hat{\alpha}^{(0)} \overline{\mathbf{A}}_{\nu}^{H} \overline{\mathbf{A}}_{\nu}+\mathbf{I}  \right)^{-1}$, $\mathbf{u}_{\overline{\mathbf{d}}_{q}(n)}^{(0)}=\hat{\beta}^{(0)} \boldsymbol{\Sigma}_{\overline{\mathbf{d}}_{q}(n)}^{(0)} \overline{\mathbf{A}}_{\tau}^{H} \mathbf{c}_{q}^H(n)$
	and
	$\boldsymbol{\Sigma}_{\overline{\mathbf{d}}_{q}(n)}^{(0)}=\left(\hat{\beta}^{(0)} \overline{\mathbf{A}}_{\tau}^{H} \overline{\mathbf{A}}_{\tau}+\mathbf{I} \right)^{-1}$, where $\hat{\gamma}^d_{(p,q)}(n)^{(0)}$  and $\hat{\gamma}^c_{(q,m)}(n)^{(0)}$ are both set to $1$. 
	}

	We also compare the performance of our proposed scheme with existing schemes. Specifically, we evaluate our proposed two-layer VBI method and the simplified two-stage MUSIC-based method using the following benchmarks: 1) \textbf{\textit{Conventional coarse FFT estimation (FFT)}} ~\cite{Sun}: The IFFT transform is applied to the channel matrix in the frequency-Doppler domain to estimate the coarse Doppler and delay; 
	2) \textbf{\textit{OTFS-based Sensing}}~\cite{OTFSfrac}: The fractional Dopplers are estimated by a difference method by taking peaks from the 2D correlation matrix;
	3)\textbf{\textit{ Two-stage VBI method}}: The VBI  method~\cite{Bayes} is adopted to calculate the row sparse matrix $\mathbf{C}(n)$ and obtain the Doppler estimates in the first stage. Then the delay estimates are obtained by VBI in the  second stage similar to Section IV;
	4) \textbf{\textit{Two-stage Expectation Maximization (EM)-VB method}}: The EM-VB method~\cite{EM} is adopted to calculate the row sparse matrix $\mathbf{C}(n)$ and obtain the Doppler estimates in the first stage. Then the delay estimates are obtained by EM-VB in the second stage similar to Section IV;
	5) \textbf{\textit{Classical VBI method with perfectly known delay or Doppler (perfect-VBI)}}~\cite{Bayes}: This method assumes that the delay or Doppler is perfectly known, allowing the VBI technique to be directly applied for Doppler or delay estimation;
	6) \textbf{\textit{Cramér-Rao bound (CRB)}}: A theoretical lower bound on the MSE for estimation performance.
	
	 In terms of complexity, the  {{FFT method}}~\cite{Sun} has a complexity of $\mathcal{O}(N^3)$, while the OTFS-based Sensing~\cite{OTFSfrac} has a complexity of $\mathcal{O}(K^2N^2)$. Considering the most demanding steps,  two-stage VBI method~\cite{Bayes} and two-stage EM-VB method~\cite{EM} both have a complexity of $\mathcal{O}((K^3+K^2Q)N^2N_{iter}'+(N^3+N^2P)NLN_{iter}'')$, where $N_{iter}'$ and $N_{iter}''$ are the iteration numbers  corresponding to the first and the second stages, respectively. The complexity of classical VBI method~\cite{Bayes} is of $\mathcal{O}((K^3+K^2Q)NLN_{iter})$, where $N_{iter}$ is the iteration number for Classical VBI method. 
	With regard to our proposed two-layer VBI method, the complexity is $\mathcal{O}((K^3+K^2Q)N^2N_{iter1}+(N^3+N^2P)QNN_{iter1}N_{iter2})$, where $N_{iter1}$ and $N_{iter2}$ are the iteration numbers of the first and the second layers in \textbf{Algorithm 1}, respectively. Compared with the above methods, our proposed method has higher computational complexity. However, it estimates the delay, fractional Doppler and integer Doppler simultaneously without mismatch issues. Moreover, since the {\it a priori} distribution assumption is more appropriate than that of the VBI-based two-stage method, its superior performance, as validated in the simulation results, justifies the increased complexity.
		Additionally, the simplified MUSIC-based two-stage method has a complexity of $\mathcal{O}(N^3K+PN^2+(K^3+K^2Q)NLN_{iter})$, which is lower than that of the two-layer VBI method. 
    Compared with the two-layer VBI algorithm, the MUSIC-based method does not require the iterative process in the first stage and reduces the  dimensions ($N \to L$) in the second stage.
This significantly reduces the overall complexity by approximately a factor of $\mathcal{O}(QN_{iter2} / L)$, assuming similar iteration counts and $L \ll N$.

\subsection{MSE Performance for the  Doppler}

\begin{figure}[t]
	\center{\includegraphics[width=0.8\columnwidth]{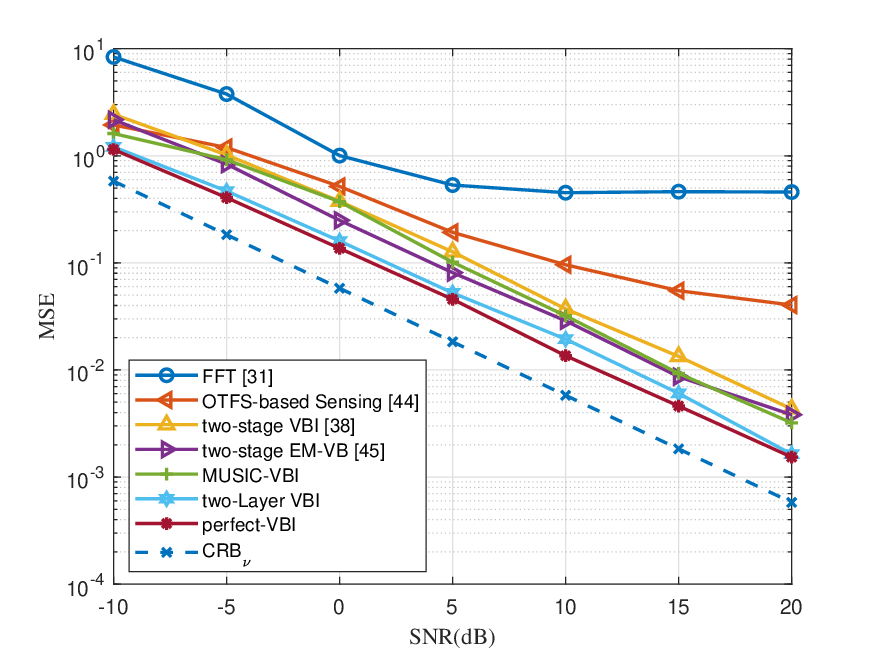}}
	\caption{The MSE performance for Doppler estimation with different algorithms against SNR.}
	\label{NMSE2}
\end{figure}

Fig. \ref{NMSE2} shows the MSE performance of our proposed methods compared with the benchmark methods. Note that the  FFT method~\cite{Sun} ignores the fractional parts of Doppler and delay, while the other benchmark methods~\cite{OTFSfrac,Bayes,EM} and our proposed algorithms account for the fractional parts of Doppler and delay. 
	It is evident that our proposed two-layer VBI outperforms the state-of-art algorithms. This owes to the capability of our designs in processing  the multiple measurements simultaneously and employing a more precise {\it a priori} distribution assumption  for the 3D-MMV model compared with the conventional VBI-based method. In addition, our proposed method is able to approach the perfect-VBI and CRB, demonstrating its high analytical performances. Furthermore, the simplified two-stage  VBI method, which is also called MUSIC-VBI, exhibits slightly larger MSE compared to the two-layer VBI method. However, it is worth noting that the MUSIC-VBI has lower complexity. 

\begin{figure}[t]
	\center{\includegraphics[width=0.8\columnwidth]{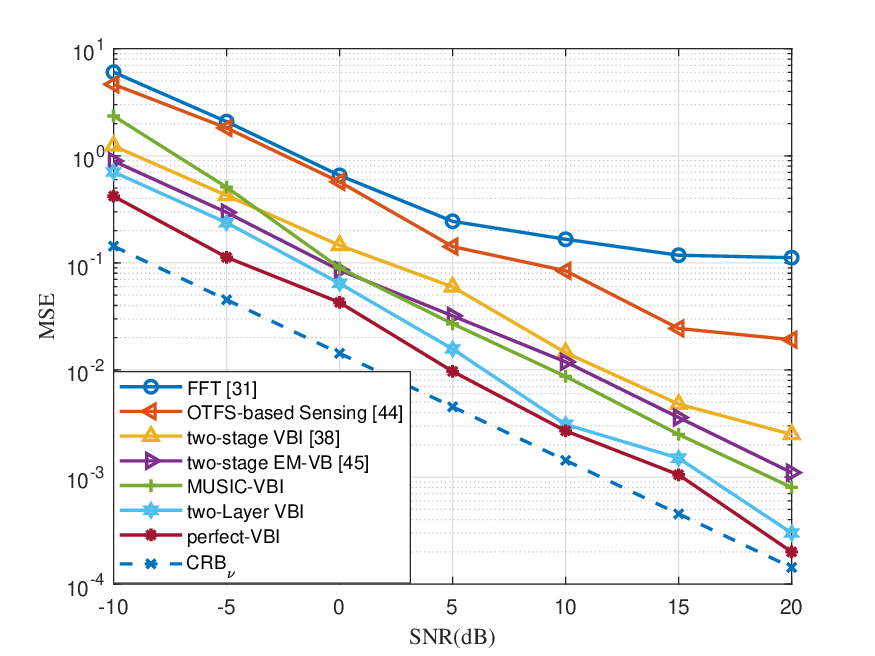}}
	\caption{{The MSE performance for Doppler estimation with different algorithms against SNR. The number of subcarriers is $N=16$ and the number of OFDM blocks is $K=32$.}}
	\label{NMSE16}
\end{figure}

 Fig. \ref{NMSE16} shows the MSE performance for the Doppler estimation of the proposed methods and CRB, where $N=16$,  $K=32$ and other parameters are set the same as in Fig. \ref{NMSE2}. We can observe that the trend of the curves is similar to that in Fig. \ref{NMSE2}, and the MSE of our proposed two-layer VBI degrades about $6$ dB at SNR = $15$ dB. It is evident that the MSE performance is slightly better than that seen in Fig. \ref{NMSE2} as the signal dimension increases.

\begin{figure}[t]
	\center{\includegraphics[width=0.8\columnwidth]{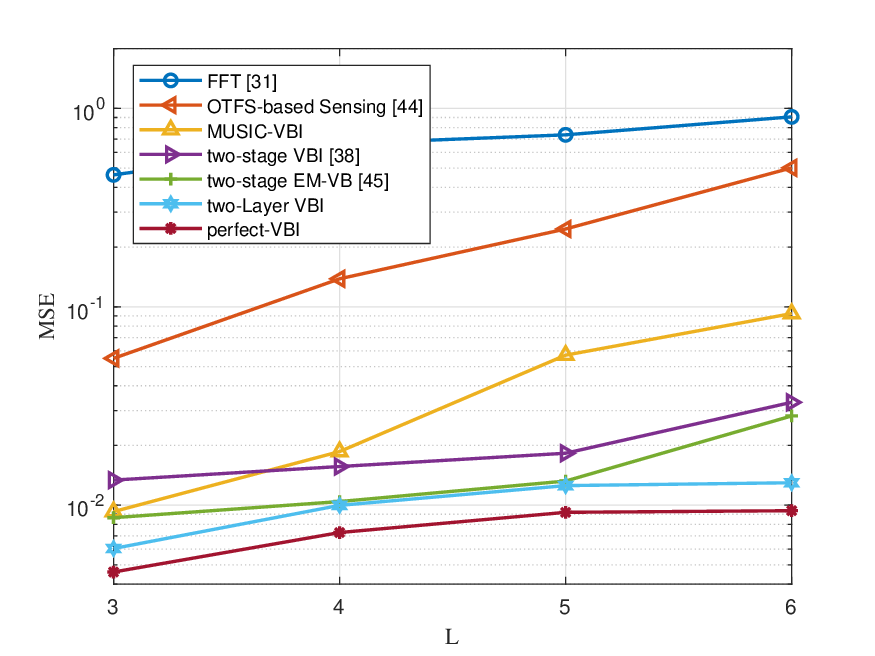}}
	\caption{The MSE performance for Doppler estimation of our proposed methods with different $L$. The SNR is fixed at $15$ dB.}
	\label{NMSEL}
\end{figure}

Fig. \ref{NMSEL} shows the MSE performance for the Doppler estimation with different numbers of paths, where the SNR is fixed at $15$ dB. As expected, the performance of all methods degrades with increasing $L$ due to the increased inter-target interference.  Despite this, our proposed two-layer VBI method consistently outperforms the existing benchmark algorithms and maintains performance close to that of the VBI method with perfectly known delay values.  Moreover, the simplified two-stage MUSIC-based VBI method exhibits more performance degradation for larger $L$, as is an inherent issue in the MUSIC algorithm \cite{MUSICproblem}. This indicates that the proposed two-layer VBI method is more robust to changes in $L$ compared with the two-stage simplified VBI method. Therefore, the latter is better suited for scenarios with fewer paths, especially under limited computational resources.

\begin{figure}[t]
	\center{\includegraphics[width=0.8\columnwidth]{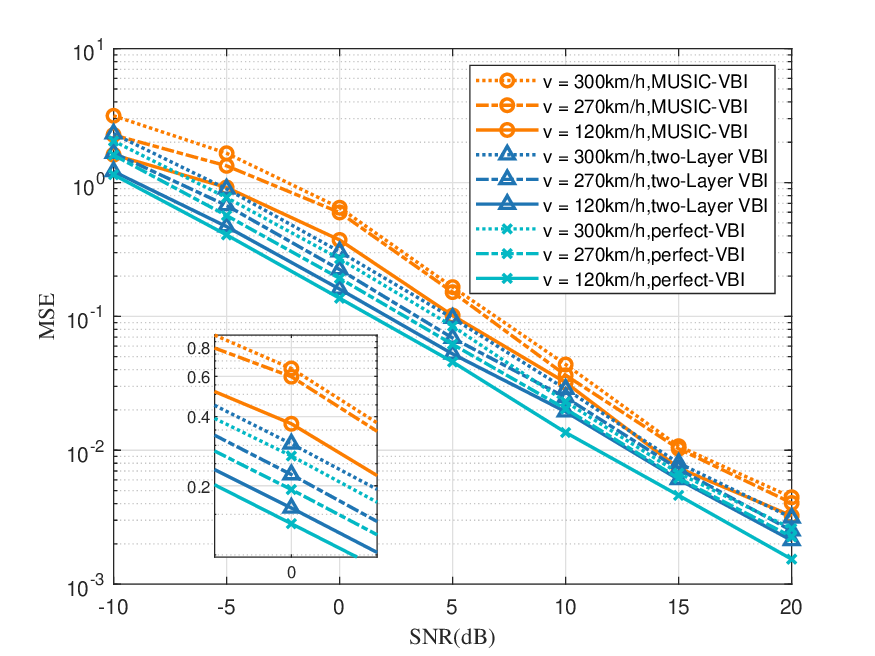}}
	\caption{The MSE performance for Doppler estimation of our proposed methods at velocities $v = 120,270,300$ km/h, corresponding to the maximum normalized  Doppler $\nu_\mathrm{max}=1.1f_0,2.5f_0,2.8f_0$, respectively.}
	\label{NMSEv}
\end{figure}

 Fig. \ref{NMSEv} shows the MSE performance for Doppler estimation against SNR at velocities of $v = 120,270,300$ km/h, corresponding to maximum normalized Dopplers of $\nu_{max}=1.1f_0,2.5f_0,2.8f_0$, respectively. It is evident that the MSE degrades as the SNR increases. Additionally, the MSE curves for both the two-stage MUSIC-based VBI method and the two-layer VBI method rise slightly as the velocity (i.e. maximum Doppler frequency) increases. However, the difference between them is minimal, which indicates that our proposed methods are robust to changes in Doppler frequency range and hence have wide applicability to diverse scenarios.

\subsection{MSE Performance for the  Delay}

\begin{figure}[t]
	\center{\includegraphics[width=0.8\columnwidth]{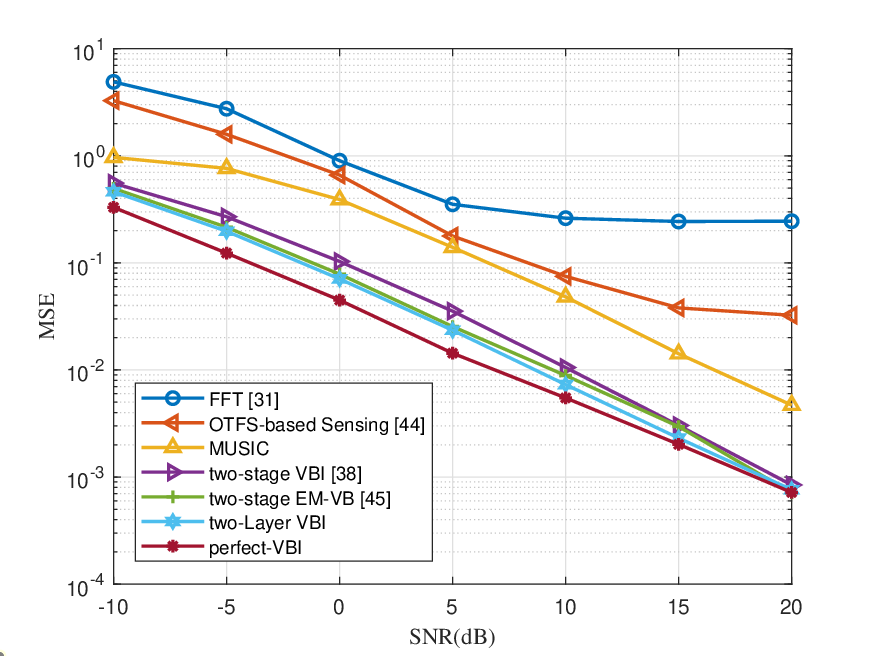}}
	\caption{The MSE performance for delay estimation with different algorithms against SNR.}
	\label{MSEtau}
\end{figure}

Fig. \ref{MSEtau} shows the MSE performance for delay using our proposed methods compared with existing methods.  It can be observed that our proposed two-layer VBI method outperforms the existing method and its MSE performance is close to that of the VBI method in the perfectly known Doppler case. We can observe that the trend of the curves is similar to that of the Doppler. Additionally, to handle multiple measurements, we use a stacking method instead of the existing summation method proposed in \cite{Sun}. Fig. \ref{MSEMU} demonstrates that our proposed stacking method outperforms the summation method, as the latter may lead to ambiguities by obscuring details during the summation process.

\begin{figure}[t]
	\center{\includegraphics[width=0.8\columnwidth]{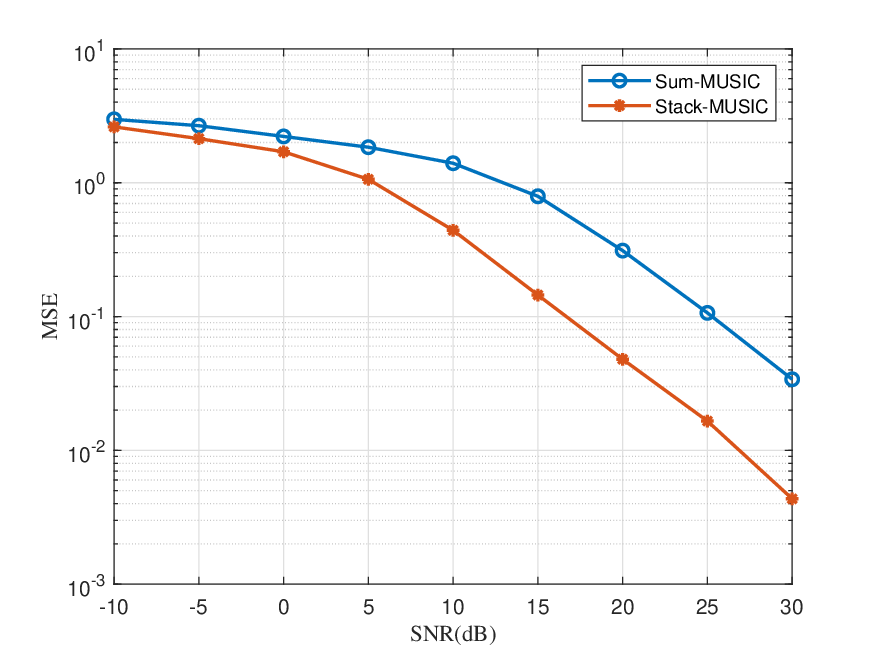}}
	\caption{The MSE performance for delay estimation of our proposed stacking method and the conventional summation method.}
	\label{MSEMU}
\end{figure}

\section{Conclusions}\label{S7}

In this paper, we proposed a two-layer VBI method to estimate the delay and the Doppler simultaneously. Firstly, the sensing parameter estimation problem was formulated as a 3D MM-SSR problem, where the positions of non-zero elements in the 3D sparse matrix correspond to the estimates of delay and Doppler. Secondly, we proposed a two-layer VBI method to avoid multiplicative expansion over multiple signal dimensions in the MM-SSR problem, where the variables in the first layer are alternately updated by exploiting the estimation results in the second layer. Thirdly, we proposed a simplified two-stage MUSIC-based VBI method to estimate the delay by MUSIC in the first stage and the Doppler by VBI in the second stage, achieving lower complexity at the cost of slight MSE performance degradation. Then the CRB was derived to characterize the theoretical lower bound. Our simulation results demonstrate that both our proposed two-layer VBI method and the simplified two-stage VBI method can achieve better MSE performance than the conventional counterparts. Furthermore, the simplified two-stage method reduces complexity with only a slight reduction in MSE performance.

\begin{appendices}

\appendix
\renewcommand{\appendixname}{APPENDICES}

\subsection {Proof of Proposition 1} \label{app0}
According to the mean field theory~\cite{bayesbase,bayesbase2}, $q(\boldsymbol{\Theta}_1)$ can be factorized as
\begin{align}
q(\boldsymbol{\Theta}_1) = \prod_{n=1}^{N} q(\boldsymbol{\Theta}_1(n)) = \prod_{n=1}^{N} q(\alpha)q(\boldsymbol{\Gamma}^c(n)) q(\mathbf{C}(n)).
\end{align}
Similarly,  $q(\boldsymbol{\Theta}_2)$ can be factorized as
\begin{align}
q(\boldsymbol{\Theta}_2) = \prod_{n=1}^{N} q(\boldsymbol{\Theta}_2(n)) = \prod_{n=1}^{N} q(\beta)q(\boldsymbol{\Gamma}^d(n)) q(\overline{\mathbf{D}}(n)).
\end{align}
Then we aim to find the optimal $q(\boldsymbol{\Theta}_1)$ and $q(\boldsymbol{\Theta}_2)$ which are closest to the MAP estimators $p\left(\boldsymbol{\Theta}_1 \mid \mathbf{Y}\right)$ and $p\left(\boldsymbol{\Theta}_2 \mid \mathbf{C}'\right)$, respectively. For the sake of simplicity, we take the calculation of $q(\boldsymbol{\Theta}_2)$  as an example. 

In order to quantify the difference between the exact {\it a posteriori} marginal distribution $p\left(\boldsymbol{\Theta}_2 \mid \mathbf{C}'\right)$ and the approximate distribution $q(\boldsymbol{\Theta}_2)$, the Kullback-Leibler (KL) divergence~\cite{bayesbase2} is introduced, which is given by
\begin{align}
\mathrm{KL}\left(q(\boldsymbol{\Theta}_2) \| p\left(\boldsymbol{\Theta}_2 \mid \mathbf{C}'\right)\right)=-\int q(\boldsymbol{\Theta}_2) \ln \frac{p\left(\boldsymbol{\Theta}_2 \mid \mathbf{C}'\right)}{q(\boldsymbol{\Theta}_2)} d \boldsymbol{\Theta}_2.
\end{align}
We aim to find the optimal probability function $q(\boldsymbol{\Theta}_2)$ closest to $p\left(\boldsymbol{\Theta}_2 \mid \mathbf{C}'\right)$, minimizing the KL divergence. Hence, the formulated problem can be expressed as follows
\begin{align} \label{theta}
q^*(\boldsymbol{\Theta}_2)=\arg \max _{q(\boldsymbol{\Theta}_2)} \int q(\boldsymbol{\Theta}_2) \ln \frac{p\left(\mathbf{C}', \boldsymbol{\Theta}_2\right)}{q(\boldsymbol{\Theta}_2)} d \boldsymbol{\Theta}_2.
\end{align}

Furthermore, (\ref{theta}) can be rewritten as the following optimization problem according to~\cite{Bayes}, which is given by
\begin{align}\label{opt}
\ln q^*\left(\boldsymbol{\Theta}_{2,z}\right)= & \left\langle\ln p\left(\mathbf{C}', \boldsymbol{\Theta}_2\right)\right\rangle_{\Pi_{i \neq z} q^*\left(\boldsymbol{\Theta}_{2,\mathbf{i}}\right)}+\text { const } \notag \\
& z=1,2,3
\end{align}
where $\boldsymbol{\Theta}_{2,z}$ denotes the $z$-th element in $\boldsymbol{\Theta}_{2}$ and $\left\langle\ln p\left(\mathbf{Y}, \boldsymbol{\Theta}_2\right)\right\rangle_{\Pi_{i \neq z} q^*\left(\boldsymbol{\Theta}_{2,\mathbf{i}}\right)}$ denotes the expection with respect to all factors in $\boldsymbol{\Theta}$ except $q^*\left(\boldsymbol{\Theta}_{2,z}\right)$. Note that the optimal probability function $\ln q^*\left(\boldsymbol{\Theta}_{2,z}\right)$ in (\ref{opt}) is dependent on other probability functions $\ln q^*\left(\boldsymbol{\Theta}_{2,j} \right), j\neq z$, it is intractable to obtain a closed-form solution. Therefore, we can obtain a stable solution instead by alternately updating the following probability functions:
\begin{align}\label{beta4}
q^{(t+1)}(\beta) \propto \exp \left(\langle\ln p(\mathbf{C}', \boldsymbol{\Theta}_2)\rangle_{q^{(t)}(\overline{\mathbf{D}}) q^{(t)}(\boldsymbol{\Gamma}^d)}\right),
\end{align}
\begin{align}\label{X4}
q^{(t+1)}(\overline{\mathbf{D}}) \propto \exp \left(\langle\ln p(\mathbf{C}', \boldsymbol{\Theta}_2)\rangle_{q^{(t+1)}(\beta) q^{(t)}(\boldsymbol{\Gamma}^d)}\right),
\end{align}
\begin{align} \label{Gamma4}
q^{(t+1)}(\boldsymbol{\Gamma}^d) \propto \exp \left(\langle\ln p(\mathbf{C}', \boldsymbol{\Theta}_2)\rangle_{q^{(t+1)}(\beta) q^{(t+1)}(\overline{\mathbf{D}})}\right),
\end{align}
where $q^{(t)}(\cdot)$ denotes the probability function of variables in the $t$-th iteration.

 \subsection{Proof of Lemma 1} \label{appA}
 
 The calculation of $q(\alpha)$ can be summarized as follows:
 \begin{align} \label{alpha1}
 &\ln q^{(i+1)}(\alpha) \notag \\  &\propto \langle\ln p(\mathbf{Y}, \boldsymbol{\Theta}_1)\rangle_{q^{(i)}(\mathbf{C}(n)) q^{(i)}(\boldsymbol{\Gamma}^c(n))} \notag \\ 
 & \propto \sum_{n=1}^{N}\left\langle\ln p\left(\mathbf{Y}(n) \mid \mathbf{C}(n), \alpha \right)\right\rangle_{q^{(i)}(\mathbf{C}(n))}+N \ln p(\alpha) \notag \\
 & \propto(a+MNQ-1) \ln (\alpha)  -\alpha b+ \notag \\ & \alpha \sum_{n=1}^N\sum_{m=1}^M\left\|\mathbf{y}_m(n)-\overline{\mathbf{A}}_{\nu} \mathbf{u}_{\mathbf{c}_m(n)}^{(t)}\right\|_{2}^{2}+\alpha \operatorname{tr}\left(\overline{\mathbf{A}}_{\nu} \boldsymbol{\Sigma}_{\mathbf{c}_m(n)}^{(i)}\left(\overline{\mathbf{A}}_{\nu}\right)^{H}\right),
 \end{align}
 where $\mathbf{u}_{\mathbf{c}_m(n)}^{(i)}=\langle\mathbf{c}_m(n)\rangle_{q^{(i)}(\mathbf{C}(n))}$ and $\boldsymbol{\Sigma}_{\mathbf{c}_m(n)}^{(i)} = \left\langle\left(\mathbf{c}_m(n)-\mathbf{u}_{\mathbf{c}_m(n)}{ }^{(i)}\right)\left(\mathbf{c}_m(n)-\mathbf{u}_{\mathbf{c}_m(n)}{ }^{(i)}\right)^{H}\right\rangle_{q^{(i)}(\mathbf{C}(n))}$.
 
 According to the expression of the Gamma distribution, we have $a^{(i+1)}_{\alpha} = a+MNQ$ and $b^{(i+1)}_{\alpha} = b+\sum_{n=1}^N\sum_{m=1}^M\left\|\mathbf{y}_m(n)-\overline{\mathbf{A}}_{\nu} \mathbf{u}_{\mathbf{c}_m(n)}^{(i)}\right\|_{2}^{2}+\operatorname{tr}\left(\overline{\mathbf{A}}_{\nu} \boldsymbol{\Sigma}_{\mathbf{c}_m(n)}^{(i)}\left(\overline{\mathbf{A}}_{\nu}\right)^{H}\right)$. Hence, $\alpha$ follows the Gamma distribution of 
 \begin{align}
 q^{(i+1)}(\alpha)=\operatorname{Gamma}\left(\alpha \mid \alpha_{\alpha}^{(i+1)}, b_{\alpha}^{(i+1)}\right).
 \end{align}
 
  \subsection{Proof of Lemma 2} \label{appB}
  
   The calculation of $q(\mathbf{C})$ can be summarized as follows:
 \begin{align} \label{C1}
 &\ln q^{(i+1)}(\mathbf{C}) \notag \\ &\propto \langle\ln p(\mathbf{Y}, \boldsymbol{\Theta}_1)\rangle_{q^{(i+1)}(\alpha) q^{(i)}(\boldsymbol{\Gamma}^c(n))}  \notag \\ & \propto \sum_{n=1}^{N}\left\langle\ln p\left(\mathbf{Y}(n) \mid \mathbf{C}(n), \alpha\right)\right\rangle_{q^{(i+1)}(\alpha)} \notag \\ &+ \sum_{n=1}^{N}\langle\ln p(\mathbf{C}(n) \mid \boldsymbol{\Gamma}^c(n))\rangle_{q^{(i)}(\boldsymbol{\Gamma}^c(n))} \notag \\ 
 & \propto-\hat{\alpha}^{(i+1)}\sum_{n=1}^N\sum_{m=1}^M\left\|\mathbf{y}_m(n)-\overline{\mathbf{A}}_{\nu}\mathbf{c}_m(n)\right\|_{2}^{2} \notag \\ &-\sum_{n=1}^N\sum_{m=1}^M\mathbf{c}_m(n)^{H} \!\! \left\langle\operatorname{diag}\{ (\hat{\boldsymbol{\gamma}}^c_m(n))^{(i)} \}\right\rangle \mathbf{c}_m(n),
 \end{align}
 where $(\hat{\boldsymbol{\gamma}}^c_m(n))^{(i)} = \left[(\hat{\gamma}_{1,m}^c(n))^{(i)}, \ldots, (\hat{\gamma}_{Q,m}^c(n))^{(i)} \right]$.
 Hence, $\mathbf{c}_m(n)$ follows the Gaussian distribution of 
 \begin{align}
 q^{(i+1)}(\mathbf{c}_m(n))=\mathcal{C N}\left(\mathbf{c}_m(n) \mid \mathbf{u}_{\mathbf{c}_m(n)}^{(i+1)}, \boldsymbol{\Sigma}_{\mathbf{c}_m(n)}^{(i+1)}\right),
 \end{align}
 where
 \begin{align} \label{Sigmac2}
 \boldsymbol{\Sigma}_{\mathbf{c}_m(n)}^{(i+1)}=\left(\hat{\alpha}^{(i+1)} \overline{\mathbf{A}}_{\nu}^{H} \overline{\mathbf{A}}_{\nu}+\left\langle\operatorname{diag}\{ (\hat{\boldsymbol{\gamma}}^c_m(n))^{(i)} \}\right\rangle \right)^{-1}
 \end{align}
 and 
 \begin{align} \label{uc2}
 \mathbf{u}_{\mathbf{c}_m(n)}^{(i+1)}=\hat{\alpha}^{(i+1)} \boldsymbol{\Sigma}_{\mathbf{c}_m(n)}^{(i+1)} \overline{\mathbf{A}}_{\nu}^{H} \mathbf{y}_m(n).
 \end{align}
 
 \subsection{Proof of Lemma 3} \label{appC}
 
 Similar to steps in (\ref{alpha1}),  $\beta$ follows the Gamma distribution of
 \begin{align}
 q^{(t+1)}(\beta)=\operatorname{Gamma}\left(\beta \mid a_{\beta}^{(t+1)}, b_{\beta}^{(t+1)}\right),
 \end{align}
 where $(t+1)$ denotes the $(t+1)$-th iteration in the second layer,  $a^{(t+1)}_{\beta} = a+NPQ$ and $b^{(t+1)}_{\beta} = b+\sum_{n=1}^N\sum_{q=1}^Q\left\|\overline{\mathbf{d}}_{q}(n)-\overline{\mathbf{A}}_{\tau} \mathbf{u}_{\overline{\mathbf{d}}_{q}(n)}^{(t)}\right\|_{2}^{2}+\operatorname{tr}\left(\overline{\mathbf{A}}_{\tau} \boldsymbol{\Sigma}_{\overline{\mathbf{d}}_{q}(n)}^{(t)}\left(\overline{\mathbf{A}}_{\tau}\right)^{H}\right)$. In the previous equation, $\overline{\mathbf{d}}_{q}(n)$ denotes the $q$-th column of $\overline{\mathbf{D}}(n)$,  $\mathbf{u}_{\overline{\mathbf{d}}_{q}(n)}^{(t)}$ and $\boldsymbol{\Sigma}_{\overline{\mathbf{d}}_{q}(n)}^{(t)}$ denote the mean  and the correlation matrix of $\overline{\mathbf{d}}_{q}(n)$ in the $t$-th iteration, respectively.  Hence, the mean of $\beta$ can be expressed as
 \begin{align} \label{beta3}
 \hat{\beta}^{(t+1)}=\langle\beta\rangle_{q^{(t+1)}(\beta)}=\frac{a_{\beta}^{(t+1)}}{b_{\beta}^{(t+1)}}.
 \end{align}
 
  \subsection{Proof of Lemma 4} \label{appD}
 
 Similarly  to steps in (\ref{C1}), {$\overline{\mathbf{d}}_{q}(n)$} follows the Gaussian distribution of
 \begin{align}
 q^{(t+1)}(\overline{\mathbf{d}}_{q}(n))=\mathcal{C N}\left(\overline{\mathbf{d}}_{q}(n) \mid \mathbf{u}_{\overline{\mathbf{d}}_{q}(n)}^{(t+1)}, \boldsymbol{\Sigma}_{\overline{\mathbf{d}}_{q}(n)}^{(t+1)}\right),
 \end{align}
 where 
 {\begin{align} \label{Sigmax2}
	\boldsymbol{\Sigma}_{\overline{\mathbf{d}}_{q}(n)}^{(t+1)}=\left(\hat{\beta}^{(t+1)} \overline{\mathbf{A}}_{\tau}^{H} \overline{\mathbf{A}}_{\tau}+\operatorname{diag}\{ (\hat{\boldsymbol{\gamma}}^d_{q}(n))^{(t)} \} \right)^{-1},
	\end{align}}
 and 
{ \begin{align} \label{ux2}
	\mathbf{u}_{\overline{\mathbf{d}}_{q}(n)}^{(t+1)}=\hat{\beta}^{(t+1)} \boldsymbol{\Sigma}_{\overline{\mathbf{d}}_{q}(n)}^{(t+1)} \overline{\mathbf{A}}_{\tau}^{H} \mathbf{c}_{q}^H(n),
	\end{align}}
 where $(\hat{\boldsymbol{\gamma}}^d_{q}(n))^{(t)} = \left[ (\hat{\gamma}_{1,q}^d(n))^{(t)}, \ldots, (\hat{\gamma}_{P,q}^d(n))^{(t)} \right]$.

   \subsection{Proof of Lemma 5} \label{appE}
 
 As for the updation of $\boldsymbol{\Gamma}^d$, we have
 \begin{align}
 \ln q^{(t+1)}(\boldsymbol{\Gamma}^d) 
 & \propto\left\langle\ln p\left(\mathbf{C}', \boldsymbol{\Theta}_2\right)\right\rangle_{q^{(t+1)}(\beta) q^{(t+1)}(\overline{\mathbf{D}})} \\ 
 & \propto\langle\ln p(\overline{\mathbf{D}} \mid \boldsymbol{\Gamma}^d)\rangle_{q^{(t+1)}(\overline{\mathbf{D}})}+\ln p(\boldsymbol{\Gamma}^d) \\ 
 & \propto \sum_{n=1}^{N}\sum_{p=1}^{P}\sum_{q=1}^{Q}({a+1}-1) \ln \gamma^d_{(p,q)}(n) \notag \\ & \!\!\!\!\!\!\!\!\!\!\!\! -\gamma^d_{(p,q)}(n) ({b+\left\langle \overline{d}_{(p,q)}(n)^{*} \overline{d}_{(p,q)}(n)\right\rangle_{q^{(t+1)}(\overline{\mathbf{D}})}})
 \end{align}
 Based on the expression of Gamma distribution, we have $a_{\gamma^d_{(p,q)}(n)}^{(t+1)} = a+1$, $b_{\gamma^d_{(p,q)}(n)}^{(t+1)} = b+\left\langle \overline{d}_{(p,q)}(n)^{*} \overline{d}_{(p,q)}(n)\right\rangle_{q^{(t+1)}(\overline{\mathbf{D}})}$, and $\gamma^d_{(p,q)}(n)$ follows the Gamma distribution of
 \begin{align}
 q^{(t+1)}\left(\gamma^d_{(p,q)}(n)\right)=\operatorname{Gamma}\left(\gamma^d_{(p,q)}(n) \mid a_{\gamma^d_{(p,q)}(n)}^{(t+1)}, b_{\gamma^d_{(p,q)}(n)}^{(t+1)} \right).
 \end{align}
 Hence, the mean of ${\gamma}^d_{(p,q)}(n)$ is 
 \begin{align} \label{gamma2}
 {\hat{\gamma}^{d}_{(p,q)}}(n)^{(t+1)}=\left\langle{\gamma}^d_{(p,q)}(n)\right\rangle_{q^{(t+1)}\left(\boldsymbol{\Gamma}^d\right)}=\frac{a_{\gamma^d_{(p,q)}(n)}^{(t+1)}}{b_{\gamma^d_{(p,q)}(n)}^{(t+1)}}.
 \end{align}

\end{appendices}

%
%

	\bibliographystyle{IEEEtran}
	\bibliography{IEEEabrv,Refference}

\begin{thebibliography}{10}
\providecommand{\url}[1]{#1}
\csname url@samestyle\endcsname
\providecommand{\newblock}{\relax}
\providecommand{\bibinfo}[2]{#2}
\providecommand{\BIBentrySTDinterwordspacing}{\spaceskip=0pt\relax}
\providecommand{\BIBentryALTinterwordstretchfactor}{4}
\providecommand{\BIBentryALTinterwordspacing}{\spaceskip=\fontdimen2\font plus
\BIBentryALTinterwordstretchfactor\fontdimen3\font minus
  \fontdimen4\font\relax}
\providecommand{\BIBforeignlanguage}[2]{{%
\expandafter\ifx\csname l@#1\endcsname\relax
\typeout{** WARNING: IEEEtran.bst: No hyphenation pattern has been}%
\typeout{** loaded for the language `#1'. Using the pattern for}%
\typeout{** the default language instead.}%
\else
\language=\csname l@#1\endcsname
\fi
#2}}
\providecommand{\BIBdecl}{\relax}
\BIBdecl

\bibitem{JCA3}
K.~Wu, J.~A. Zhang, Z.~Ni, X.~Huang, Y.~J. Guo, and S.~Chen, ``Joint
  communications and sensing employing optimized {MIMO}-{OFDM} signals,''
  \emph{IEEE Internet Things J.}, vol.~11, no.~6, pp. 10\,368--10\,383, March
  2024.

\bibitem{JCA4}
S.~D. Liyanaarachchi, T.~Riihonen, C.~B. Barneto, and M.~Valkama, ``Joint
  {MIMO} communications and sensing with hybrid beamforming architecture and
  {OFDM} waveform optimization,'' \emph{{IEEE} Trans. Wireless Commun.},
  vol.~23, no.~2, pp. 1565--1580, Feb. 2024.

\bibitem{JCA5}
X.~Chen, Z.~Feng, Z.~Wei, P.~Zhang, and X.~Yuan, ``Code-division {OFDM} joint
  communication and sensing system for {6G} machine-type communication,''
  \emph{IEEE Internet Things J.}, vol.~8, no.~15, pp. 12\,093--12\,105, Aug.
  2021.

\bibitem{JCA6}
F.~Liu, C.~Masouros, A.~P. Petropulu, H.~Griffiths, and L.~Hanzo, ``Joint radar
  and communication design: Applications, state-of-the-art, and the road
  ahead,'' \emph{{IEEE} Trans. Commun.}, vol.~68, no.~6, pp. 3834--3862, Jun.
  2020.

\bibitem{JCAS1}
J.~A. Zhang, F.~Liu, C.~Masouros, R.~W. Heath, Z.~Feng, L.~Zheng, and
  A.~Petropulu, ``An overview of signal processing techniques for joint
  communication and radar sensing,'' \emph{{IEEE} J. Sel. Topics Signal
  Process.}, vol.~15, no.~6, pp. 1295--1315, Nov. 2021.

\bibitem{ISAC1}
Z.~Wei, F.~Liu, C.~Masouros, N.~Su, and A.~P. Petropulu, ``Toward
  multi-functional {6G} wireless networks: Integrating sensing, communication,
  and security,'' \emph{{IEEE} Commun. Mag.}, vol.~60, no.~4, pp. 65--71, April
  2022.

\bibitem{ISACBayes3}
Q.~Tao, C.~Huang, and X.~Chen, ``Integrated sensing and communication for
  symbiotic radio with multiple {IoT} devices,'' \emph{{IEEE} Commun. Lett.},
  vol.~28, no.~8, pp. 1820--1824, Aug. 2024.

\bibitem{ISACBayes4}
Z.~Hu, A.~Liu, Y.~Wan, T.~Q.~S. Quek, and M.-J. Zhao, ``Two-stage multiband
  {Wi-Fi} sensing for {ISAC} via stochastic particle-based variational bayesian
  inference,'' in \emph{GLOBECOM 2023 - 2023 IEEE Global Communications
  Conference}, 2023, pp. 5617--5622.

\bibitem{ISACByesa1}
X.~Gan, C.~Huang, Z.~Yang, C.~Zhong, X.~Chen, Z.~Zhang, Q.~Guo, C.~Yuen, and
  M.~Debbah, ``Bayesian learning for double-{RIS} aided {ISAC} systems with
  superimposed pilots and data,'' \emph{{IEEE} J. Sel. Topics Signal Process.},
  vol.~18, no.~5, pp. 766--781, July 2024.

\bibitem{ISACByesa2}
K.~Chen and C.~Qi, ``Joint sparse bayesian learning for channel estimation in
  {ISAC},'' \emph{{IEEE} Commun. Lett.}, vol.~28, no.~8, pp. 1825--1829, Aug.
  2024.

\bibitem{JCA2}
J.~A. Zhang, M.~L. Rahman, K.~Wu, X.~Huang, Y.~J. Guo, S.~Chen, and J.~Yuan,
  ``Enabling joint communication and radar sensing in mobile networks—a
  survey,'' \emph{IEEE Commun. Surveys and Tutorials}, vol.~24, no.~1, pp.
  306--345, Oct. 2022.

\bibitem{OFDM2}
X.~Cai and G.~B. Giannakis, ``Bounding performance and suppressing intercarrier
  interference in wireless mobile {OFDM},'' \emph{{IEEE} Trans. Commun.},
  vol.~51, no.~12, pp. 2047--2056, 2003.

\bibitem{OFDM3}
Y.~S. Cho, J.~Kim, W.~Y. Yang, and C.~G. Kang, \emph{{MIMO}-{OFDM} {Wireless}
  {Communications} {With} {MATLAB}}.\hskip 1em plus 0.5em minus 0.4em\relax
  Singapore: Wiley (Asia) Pte. Ltd., 2010.

\bibitem{OFDM}
H.~Sari, G.~Karam, and I.~Jeanclaude, ``Transmission techniques for digital
  terrestrial {TV} broadcasting,'' \emph{{IEEE} Commun. Mag.}, vol.~33, no.~2,
  pp. 100--109, 1995.

\bibitem{OTFSz}
X.~Zhang, C.~Liu, W.~Yuan, J.~A. Zhang, and D.~W.~K. Ng, ``Sparse prior-guided
  deep learning for {OTFS} channel estimation,'' \emph{{IEEE} Trans. Veh.
  Technol.}, vol.~73, no.~12, pp. 19\,913--19\,918, Dec. 2024.

\bibitem{OTFS1}
R.~Hadani and A.~Monk, ``{OTFS:} a new generation of modulation addressing the
  challenges of {5G},'' \emph{arXiv preprint arXiv:1802.02623}, 2018.

\bibitem{OTFS2}
R.~Hadani, S.~Rakib, M.~Tsatsanis, A.~Monk, A.~J. Goldsmith, A.~F. Molisch, and
  R.~Calderbank, ``Orthogonal time frequency space modulation,'' in \emph{Proc.
  IEEE WCNC}, 2017, pp. 1--6.

\bibitem{OTFS3}
Z.~Wei, W.~Yuan, S.~Li, J.~Yuan, G.~Bharatula, R.~Hadani, and L.~Hanzo,
  ``Orthogonal time-frequency space modulation: A promising next-generation
  waveform,'' \emph{IEEE Wireless Commun.}, vol.~28, no.~4, pp. 136--144, Aug.
  2021.

\bibitem{III}
W.~Yuan, Z.~Wei, S.~Li, R.~Schober, and G.~Caire, ``Orthogonal time frequency
  space modulation—part {III}: {ISAC} and potential applications,''
  \emph{{IEEE} Commun. Lett.}, vol.~27, no.~1, pp. 14--18, Jan. 2023.

\bibitem{YUAN1}
W.~Yuan, Z.~Wei, S.~Li, J.~Yuan, and D.~W.~K. Ng, ``Integrated sensing and
  communication-assisted orthogonal time frequency space transmission for
  vehicular networks,'' \emph{{IEEE} J. Sel. Topics Signal Process.}, vol.~15,
  no.~6, pp. 1515--1528, Nov. 2021.

\bibitem{NOMA3D}
L.~Xiang, K.~Xu, J.~Hu, C.~Masouros, and K.~Yang, ``Robust {NOMA}-assisted
  {OTFS}-{ISAC} network design with 3-{D} motion prediction topology,''
  \emph{IEEE Internet Things J.}, vol.~11, no.~9, pp. 15\,909--15\,918, May
  2024.

\bibitem{BIAMP}
X.~Yang, H.~Li, Q.~Guo, J.~A. Zhang, X.~Huang, and Z.~Cheng, ``Sensing aided
  uplink transmission in {OTFS} {ISAC} with joint parameter association,
  channel estimation and signal detection,'' \emph{{IEEE} Trans. Veh.
  Technol.}, vol.~73, no.~6, pp. 9109--9114, Jun. 2024.

\bibitem{frac1}
Z.~Wei, W.~Yuan, S.~Li, J.~Yuan, and D.~W.~K. Ng, ``Off-grid channel estimation
  with sparse {Bayesian} learning for {OTFS} systems,'' \emph{{IEEE} Trans.
  Wireless Commun.}, vol.~21, no.~9, pp. 7407--7426, Sep. 2022.

\bibitem{frac2}
Y.~Shan, F.~Wang, Y.~Hao, J.~Yuan, J.~Hua, and Y.~Xin, ``Off-grid channel
  estimation using grid evolution for {OTFS} systems,'' \emph{{IEEE} Trans.
  Wireless Commun.}, pp. 1--1, 2024.

\bibitem{SS}
S.~Li, W.~Yuan, C.~Liu, Z.~Wei, J.~Yuan, B.~Bai, and D.~W.~K. Ng, ``A novel
  {ISAC} transmission framework based on spatially-spread orthogonal time
  frequency space modulation,'' \emph{{IEEE} J. Sel. Areas Commun.}, vol.~40,
  no.~6, pp. 1854--1872, Jun. 2022.

\bibitem{DFT}
Y.~Wu, C.~Han, and Z.~Chen, ``{DFT}-spread orthogonal time frequency space
  system with superimposed pilots for terahertz integrated sensing and
  communication,'' \emph{{IEEE} Trans. Wireless Commun.}, vol.~22, no.~11, pp.
  7361--7376, Nov. 2023.

\bibitem{IIOT}
K.~Wu, J.~A. Zhang, X.~Huang, and Y.~J. Guo, ``{OTFS}-based joint communication
  and sensing for future industrial {IoT},'' \emph{IEEE Internet Things J.},
  vol.~10, no.~3, pp. 1973--1989, Feb. 2023.

\bibitem{REFINE}
Y.~Shi and Y.~Huang, ``Integrated sensing and communication-assisted user state
  refinement for {OTFS} systems,'' \emph{{IEEE} Trans. Wireless Commun.},
  vol.~23, no.~2, pp. 922--936, Feb. 2024.

\bibitem{ZHY1}
H.~Zhang, X.~Huang, and J.~A. Zhang, ``Adaptive transmission with
  frequency-domain precoding and linear equalization over fast fading
  channels,'' \emph{{IEEE} Trans. Wireless Commun.}, vol.~20, no.~11, pp.
  7420--7430, Nov. 2021.

\bibitem{ZHY2}
------, ``Low-overhead {OTFS} transmission with frequency or time domain
  channel estimation,'' \emph{{IEEE} Trans. Veh. Technol.}, vol.~73, no.~1, pp.
  799--811, Jan. 2024.

\bibitem{Sun}
Y.~Sun, J.~A. Zhang, K.~Wu, and R.~P. Liu, ``Frequency-domain sensing in
  time-varying channels,'' \emph{IEEE Wireless Commun. Lett.}, vol.~12, no.~1,
  pp. 16--20, Jan. 2023.

\bibitem{bayesbase}
D.~P. Wipf and B.~D. Rao, ``Sparse {Bayesian} learning for basis selection,''
  \emph{{IEEE} Trans. Signal Process.}, vol.~52, no.~8, pp. 2153--2164, 2004.

\bibitem{bayesbase2}
M.~E. Tipping, ``Sparse {Bayesian} learning and the relevance vector machine,''
  \emph{J. Mach. Learn. Res.}, vol.~1, no. Jun., pp. 211--244, 2001.

\bibitem{bayestutorial}
C.~W. Fox and S.~J. Roberts, ``A tutorial on variational {Bayesian}
  inference,'' \emph{Artificial Intelligence Rev.}, vol.~38, no.~2, pp. 85--95,
  2012.

\bibitem{Gaussian}
G.~Ozcan, M.~C. Gursoy, and S.~Gezici, ``Error rate analysis of cognitive radio
  transmissions with imperfect channel sensing,'' \emph{{IEEE} Trans. Wireless
  Commun.}, vol.~13, no.~3, pp. 1642--1655, March 2014.

\bibitem{Gaussian2}
A.~Mishra and A.~K. Jagannatham, ``Sbl-based glrt for spectrum sensing in
  ofdma-based cognitive radio networks,'' \emph{{IEEE} Commun. Lett.}, vol.~20,
  no.~7, pp. 1433--1436, July 2016.

\bibitem{WXY}
X.~Wang, W.~Shen, C.~Xing, J.~An, and L.~Hanzo, ``Joint {Bayesian} channel
  estimation and data detection for {OTFS} systems in {LEO} satellite
  communications,'' \emph{{IEEE} Trans. Commun.}, vol.~70, no.~7, pp.
  4386--4399, Jul. 2022.

\bibitem{Bayes}
D.~G. Tzikas, A.~C. Likas, and N.~P. Galatsanos, ``The variational
  approximation for {Bayesian} inference,'' \emph{{IEEE} Signal Process. Mag.},
  vol.~25, no.~6, pp. 131--146, 2008.

\bibitem{MUSIC2}
X.~Zhang, L.~Xu, L.~Xu, and D.~Xu, ``Direction of departure ({DOD}) and
  direction of arrival ({DOA}) estimation in {MIMO} radar with
  reduced-dimension {MUSIC},'' \emph{{IEEE} Commun. Lett.}, vol.~14, no.~12,
  pp. 1161--1163, Dec. 2010.

\bibitem{MUSIC4}
F.~Yan, M.~Jin, and X.~Qiao, ``Low-complexity {DOA} estimation based on
  compressed {MUSIC} and its performance analysis,'' \emph{{IEEE} Trans. Signal
  Process.}, vol.~61, no.~8, pp. 1915--1930, April 2013.

\bibitem{MUSIC5}
P.~Vallet, X.~Mestre, and P.~Loubaton, ``Performance analysis of an improved
  {MUSIC} {DoA} estimator,'' \emph{{IEEE} Trans. Signal Process.}, vol.~63,
  no.~23, pp. 6407--6422, Dec. 2015.

\bibitem{MUSIC}
X.~Chen, Z.~Feng, Z.~Wei, X.~Yuan, P.~Zhang, J.~A. Zhang, and H.~Yang,
  ``Multiple signal classification based joint communication and sensing
  system,'' \emph{{IEEE} Trans. Wireless Commun.}, vol.~22, no.~10, pp.
  6504--6517, Oct. 2023.

\bibitem{estimate}
S.~M. Kay, \emph{Fundamentals of Statistical Signal Processing: Estimation
  Theory}.\hskip 1em plus 0.5em minus 0.4em\relax Upper Saddle River, NJ, USA:
  Prentice Hall, 1993.

\bibitem{OTFSfrac}
K.~Zhang, W.~Yuan, S.~Li, F.~Liu, F.~Gao, P.~Fan, and Y.~Cai, ``Radar sensing
  via {OTFS} signaling: A delay {Doppler} signal processing perspective,'' in
  \emph{ICC 2023 - IEEE International Conference on Communications}, 2023, pp.
  6429--6434.

\bibitem{EM}
J.~Dai, X.~Bao, W.~Xu, and C.~Chang, \emph{{IEEE} Signal Process. Lett.}

\bibitem{MUSICproblem}
\BIBentryALTinterwordspacing
H.~Tang, \emph{{DOA} Estimation Based on {MUSIC} Algorithm}, {Accessed}: Nov.
  28, 2017. [Online]. Available:
  \url{https://www.divaportal.org/smash/get/diva2:724272/FULLTEXT01.pdf}
\BIBentrySTDinterwordspacing

\end{thebibliography}
	
\end{document}